\documentclass[onecolumn]{article}

\usepackage[english]{babel}

\usepackage[letterpaper,top=2cm,bottom=2cm,left=2cm,right=2cm,marginparwidth=1.75cm]{geometry}

\usepackage{footmisc}
\usepackage{appendix}
\usepackage{amssymb}
\usepackage{amsmath}
\usepackage{appendix}
\usepackage{float}
\usepackage{graphbox}
\usepackage{authblk}
\usepackage[colorlinks=true, allcolors=blue]{hyperref}
\usepackage{subcaption}
\usepackage{indentfirst}
\usepackage{graphicx}
\usepackage{comment}
\usepackage{multirow}
\usepackage{makecell}
\usepackage{xcolor}
\usepackage[figurename=Fig.]{caption}
\usepackage{cite}
\usepackage{amsfonts}
\usepackage{algorithm}
\usepackage{algorithmic}
\usepackage{rotating} 
\usepackage{booktabs}

\title{Neural quantum support vector data description for one-class classification}
\author[1]{\normalsize Changjae Im}
\author[1]{\normalsize Hyeondo Oh}
\author[1,2,3,*]{\normalsize Daniel K. Park}
\affil[1]{\small \textit{Department of Statistics and Data Science, Yonsei University, Seoul, Republic of Korea}}
\affil[2]{\small \textit{Department of Applied Statistics, Yonsei University, Seoul, Republic of Korea}}
\affil[3]{\small \textit{Department of Quantum Information, Yonsei University, Seoul, Republic of Korea}}

\date{}

\begin{document}

\maketitle
{\let\thefootnote\relax\footnotetext{\small * : corresponding author}}
{\let\thefootnote\relax\footnotetext{\small Email addresses : dkd.park@yonsei.ac.kr}}

\begin{abstract}
One-class classification (OCC) is a fundamental problem in machine learning with numerous applications, such as anomaly detection and quality control. With the increasing complexity and dimensionality of modern datasets, there is a growing demand for advanced OCC techniques with better expressivity and efficiency. We introduce Neural Quantum Support Vector Data Description (NQSVDD), a classical–quantum hybrid framework for OCC that performs end-to-end optimized hierarchical representation learning. NQSVDD integrates a classical neural network with trainable quantum data encoding and a variational quantum circuit, enabling the model to learn nonlinear feature transformations tailored to the OCC objective. The hybrid architecture maps input data into an intermediate high-dimensional feature space and subsequently projects it into a compact latent space defined through quantum measurements. Importantly, both the feature embedding and the latent representation are jointly optimized such that normal data form a compact cluster, for which a minimum-volume enclosing hypersphere provides an effective decision boundary. Experimental evaluations on benchmark datasets demonstrate that NQSVDD achieves competitive or superior AUC performance compared to classical Deep SVDD and quantum baselines, while maintaining parameter efficiency and robustness under realistic noise conditions.
\end{abstract}

\section{Introduction}
Quantum machine learning (QML) has demonstrated promising capabilities in data analysis and pattern recognition~\cite{Cerezo_2022,Schuld_2014,Biamonte_2017}. In the field of binary classification, quantum computing techniques offer the potential to exceed classical algorithms in terms of efficiency, expressivity, and accuracy~\cite{Rebentrost_2014,Benedetti_2019,Cerezo_2021_vqa,Abbas_2021}. However, the task of one-class classification (OCC)~\cite{MOYA1996463,Khan_2014,perera2021oneclassclassificationsurvey,chalapathy2019deeplearninganomalydetection,NIPS1999_8725fb77,Tax2004,pmlr-v80-ruff18a} remains particularly challenging, as the training data consists solely of examples from a single class. Without access to contrasting or negative samples, OCC models must tightly characterize the target class in order to detect deviations, typically in a semi-supervised or unsupervised setting. Despite these challenges, OCC has a wide range of real-world applications, such as finance~\cite{phua2010comprehensive,li2012identifying,jeragh2018combining}, cybersecurity~\cite{GARCIATEODORO200918}, bioinformatics~\cite{feher2014cell,min2017deep}, and particle physics~\cite{Fraser2022}.

Several studies have proposed QML for OCC~\cite{Liu_2018,Kottmann_2021,Sakhnenko_2022,Ngairangbam_2022,Park_2023,Oh_2024}, yet many face limitations in practice. Some approaches require complex subroutines beyond the capabilities of current quantum hardware~\cite{Oh_2024}, while others focus more on data compression than effective anomaly detection~\cite{Sakhnenko_2022,Ngairangbam_2022,Park_2023}. To address these challenges, we propose \textit{Neural Quantum Support Vector Data Description (NQSVDD)}, a classical-quantum hybrid model designed for OCC of classical data on noisy intermediate-scale quantum (NISQ) devices. NQSVDD integrates a classical neural network with a quantum feature map for early-stage feature extraction, enabling rich expressive representations within a high-dimensional Hilbert space. Also, the hybrid approach balances computational complexity, alleviating the need for higher-complexity quantum processing~\cite{Cerezo_2022}. Quantum feature mapping consists of alternations of the embedding layers that converts input data into quantum states and the parameterized layers with learnable parameters, increasing the representation capability. This is followed by a variational quantum circuit (VQC) with local measurements, such as a quantum convolutional neural network (QCNN)~\cite{cong_QCNN}, for identifying a minimum-volume hypersphere enclosing the training data in a latent space to define the decision boundary. While various strategies can be employed for learning parameters, it is basically possible to update the parameters of the initial classical neural network, the parameterized layers of the quantum feature mapping, and the VQC simultaneously. This hybrid approach, which jointly learns both quantum and classical neural networks, has a potential to show OCC performance over classical algorithms in NISQ devices with a limited number of available qubits.

Experimental results on the MNIST~\cite{726791}, Fashion-MNIST~\cite{xiao2017fashionmnist} image datasets, Credit Card Transaction dataset~\cite{kagglecredit}, and Network Intrusion (CIC-IDS2017) dataset~\cite{kagglenetwork}, implemented using Pennylane~\cite{bergholm2020pennylane}, demonstrates that NQSVDD outperforms both classical and quantum baselines, achieving higher accuracy with fewer trainable parameters. For the MNIST and Fashion-MNIST datasets, each consists of ten different classes. We treat one of the classes as the target class and the remaining as outliers, resulting in ten independent OCC experiment setups for each dataset. Only samples from the respective target class are used to optimize the hypersphere. Performance is evaluated on the test dataset, which consists of balanced number of target and outlier samples, using the area under the receiver operating characteristic (ROC) curve (AUC). In most cases, NQSVDD outperforms QSVDD, a solely-quantum method with amplitude encoding, and the classical DSVDD, in both ideal and noisy environments. Notably, NQSVDD achieves better classification performance than QSVDD, with a more NISQ-compatible architecture. Furthermore, NQSVDD outperforms DSVDD, despite learning fewer number of parameters than DSVDD. These results indicate that NQSVDD is an effective and efficient OCC algorithm, well-suited for NISQ era, as it combines classical robustness with quantum feature expressivity while maintaining shallow circuit depth and low qubit requirements.

\section{Background}
\subsection{One-class classification}

Let $\mathcal{X}\subseteq\mathbb{R}^{d}$ denote the input space and $\mathcal{Y}=\{0,1,...,K-1\}$ be the label space for $K$ classes. Given a labeled training dataset $\mathcal{D}_{K}=\{(\boldsymbol{x}_{i},y_{i})\}_{i=1}^{m}\subset\mathcal{X}\times\mathcal{Y}$, the goal of classification is to learn a function $f:\mathcal{X}\rightarrow\mathcal{Y}$ that can accurately predict the class label of an unseen input $\tilde{\boldsymbol{x}}\in\mathcal{X}$.

One-class classification (OCC) is an extreme case of classification, where data points from only a single target class are present during training. The absence of prior knowledge about samples outside the target class makes the task more challenging compared to general multi-class classification. Given a training dataset $\mathcal{D}_{K}$ with $K=1$, which is reduced to 
$\mathcal{D}_{1}=\{\boldsymbol{x}_{1},...,\boldsymbol{x}_{m}\}\subset\mathcal{X}_{\text{target}}\subseteq\mathcal{X}$,
the objective of OCC is to learn a decision function that distinguishes whether an unseen data point belongs to the distribution of the target class, or not. More specifically, suppose $f:\mathcal{X}\rightarrow\mathbb{R}$ be a decision function that measures the similarity of an input $\tilde{\boldsymbol{x}}\in\mathcal{X}$ to the representation of the target class. For a predefined threshold $b$, if $f(\tilde{\boldsymbol{x}};\mathcal{D}_{1}))\leq b$, $\tilde{\boldsymbol{x}}$ is classified as the target object. Conversely, if $f(\tilde{\boldsymbol{x}};\mathcal{D}_{1}))> b$, then $\tilde{\boldsymbol{x}}$ is identified as an outlier.
\begin{equation}\label{eq:occ_decision}
    h(f(\tilde{\boldsymbol{x}};\mathcal{D}_{1}))=
    \begin{cases}
        1 & \text{if}\;f(\tilde{\boldsymbol{x}};\mathcal{D}_{1})\leq b\\
        0 & \text{if}\;f(\tilde{\boldsymbol{x}};\mathcal{D}_{1})>b.
    \end{cases}
\end{equation}
The compact feature representation of target instances and distinctive representations between different classes are two desirable properties of OCC~\cite{perera2021oneclassclassificationsurvey}. In other words, the compactness and descriptiveness of the decision function is directly related to the performance of OCC.

\subsection{Deep support vector data description}

Support vector data description (SVDD)~\cite{Tax2004} is a kernel-based method for OCC, which aims to make a description of a training dataset by constructing a minimum volume hypersphere in feature space that encloses the majority of the training data. This hypersphere provides a boundary for distinguishing target data from outliers. 

Let $k:\mathcal{X}\times\mathcal{X}\rightarrow[0,\infty)$ be a positive semi-definite (PSD) kernel, $\mathcal{F}_{k}$ be its associated reproducing kernel Hilbert space (RKHS), and $\phi_{k}:\mathcal{X}\rightarrow\mathcal{F}_{k}$ be its corresponding feature mapping, such that $k(\boldsymbol{x}, \tilde{\boldsymbol{x}})=\langle\phi_{k}(\boldsymbol{x}),\phi_{k}(\tilde{\boldsymbol{x}})\rangle_{\mathcal{F}_{k}}$ for all $\boldsymbol{x},\tilde{\boldsymbol{x}}\in\mathcal{X}$. Then, given a dataset $\mathcal{D}=\{\boldsymbol{x}_{i}\}_{i=1}^{m}$ with $\boldsymbol{x}_{i}\in\mathcal{X}$, for a center $\boldsymbol{c}\in\mathcal{F}_{k}$ and a radius $R>0$ that uniquely define a hypersphere, the primal optimization problem of SVDD is formulated as
\begin{equation}\label{eq:svdd}
    \begin{aligned}
        &\min_{R, \boldsymbol{c}, \boldsymbol{\xi}}\quad R^{2}+\frac{1}{\nu m}\sum_{i}\xi_{i}\\
        &\text{s.t.}\quad\|\phi_{k}(\boldsymbol{x}_{i})-\boldsymbol{c}\|^{2}_{\mathcal{F}_{k}}\leq R^{2}+\xi_{i},\;\xi_{i}\geq0,
    \end{aligned}
\end{equation}
where $\xi_{i}\geq0$ are slack variables allowing for soft violations of the boundary, and a hyperparameter $\nu\in(0,1]$ controls the trade-off between the volume of the sphere and penalties $\xi_{i}$. 

The SVDD method suffers from the necessity to perform explicit feature engineering, at least quadratic computational scaling in the number of samples, and high memory usage to store support vectors. These limitations motivate deep support vector data description (DSVDD)~\cite{pmlr-v80-ruff18a}, which is a deep learning approach of SVDD. Instead of explicitly choosing the kernel function, DSVDD learns a compact feature representation of the data by training a neural network to map the data into a hypersphere of minimum volume in the latent space. The loss function of DSVDD for finding the optimal feature space and hypersphere is defined as
\begin{equation}\label{eq:dsvdd_loss}
\mathcal{L}_{c}({\mathcal{W}})=\frac{1}{n}\sum_{i=1}^{m}\|\varphi_{c}(\boldsymbol{x}_{i};\mathcal{W})-\boldsymbol{c}\|^{2}+\frac{\lambda}{2}\sum_{\ell=1}^{L}\|\boldsymbol{W}^{\ell}\|_{F}^{2}.
\end{equation}
Here, for some input space $\mathcal{X}\subseteq\mathbb{R}^{d}$ and output space $\mathcal{F}\in\mathbb{R}^{d'}$ with $d'<d$, $\varphi_{c}(\cdot;\mathcal{W}):\mathcal{X}\rightarrow\mathcal{F}$ is a neural network with $L\in\mathbb{N}$ hidden layers and a set of trainable parameters $\mathcal{W}=\{\boldsymbol{W}^{1},...,\boldsymbol{W}^{L}\}$ where $\boldsymbol{W}^{\ell}$ are the parameters of layer $\ell\in\{1,...,L\}$. The first term penalizes the mean distance of all network representation $\varphi_{c}(\boldsymbol{x}_{i};\mathcal{W})$ to the center $\boldsymbol{c}\in\mathcal{F}$. In order to avoid the trivial all-zero-weight solution, leading to the hypersphere collapse, the predefined center $\boldsymbol{c}$ is fixed to be nonzero. Also, the network should exclude the bias term and use unbounded activation functions such as the ReLU. The second term is a weight decay regularizer on the network parameter $\mathcal{W}$ with hyperparameter $\lambda>0$, where $\|\cdot\|_{F}$ denotes the Frobenius norm. Given a test data point $\tilde{\boldsymbol{x}}\in\mathcal{X}$, the decision function is expressed as
\begin{equation}\label{eq:dsvdd_decision}
f(\tilde{\boldsymbol{x}})=\|\varphi_{c}(\tilde{\boldsymbol{x}};\mathcal{W}^{*})-\boldsymbol{c}\|^{2}-R^{2*},
\end{equation}
where $\mathcal{W}^{*}$ are the trained parameters and $R^{2*}=\max_{i}\|\varphi_{c}(\boldsymbol{x}_{i};\mathcal{W}^{*})-\boldsymbol{c}\|^{2}$ is the final radius.

\subsection{Quantum support vector data description}

Inspired by DSVDD that introduces classical neural network, quantum support vector data description (QSVDD)~\cite{Oh_2024} leverages shallow-depth variational QML algorithm, addressing the limitation of prior methods that are impractical for NISQ devices or not directly designed or anomaly detection. In order to process the classical data on a quantum computer, quantum data encoding $\Psi:\mathcal{X}\rightarrow\mathcal{H}$ maps input data $\boldsymbol{x}\in\mathcal{X}$ to quantum state $|\psi(\boldsymbol{x})\rangle\in\mathbb{C}^{2^{n}}$ in the Hilbert space $\mathcal{H}$ whose dimension grows exponentially with the number of qubits $n$. This can be done by applying a unitary transformation $U_{\psi}(\boldsymbol{x})$ to an initial state $|0\rangle^{\otimes n}$, represented as $\boldsymbol{x}\in\mathcal{X}\mapsto|\psi(\boldsymbol{x})\rangle=U_{\psi}(\boldsymbol{x})|0\rangle^{\otimes n}\in\mathcal{H}$. While there are various kinds of quantum feature map, QSVDD focuses on amplitude encoding, which encodes $N$-dimensional vector $\boldsymbol{x}=(x_{1},...x_{N})^\top$ into a quantum state
\begin{equation}\label{eq:amplitudeencoding}
|\psi(\boldsymbol{x})\rangle=\frac{1}{\|\boldsymbol{x}\|}\sum_{i=1}^{N}x_{i}|i\rangle,
\end{equation}
where $N=2^{n}$, and $|i\rangle$ is the $i$th computational basis state.

After preparing the encoded quantum state, variational quantum circuit (VQC) $U(\boldsymbol{\theta})$, parameterized by $\boldsymbol{\theta}$, learns the data representation in the quantum Hilbert space. QSVDD chooses the quantum convolutional neural network (QCNN), which iterates quantum convolution and pooling layers until the remaining system has sufficiently small number of active qubits. The final $n_{f}$-qubit quantum state, which is fully characterized by $4^{n_{f}}-1$ parameters, is then mapped into the final latent space through a quantum measurement expectation function $g:\mathbb{C}^{2^{n}}\rightarrow\mathbb{R}^{d'}$. The dimension of the latent space $d'$ is determined by the selection of Pauli observables $P_{i}$ from the set of $n_{f}$-qubit Pauli observables. The feature mapping of QSVDD is expressed as
\begin{equation}\label{eq:qsvdd}
\varphi_{q}(\boldsymbol{x}, \boldsymbol{\theta})=g(U(\boldsymbol{\theta})|\psi(\boldsymbol{x})\rangle).
\end{equation}
Given a training target data samples $\{\boldsymbol{x}_{i}\}_{i=1}^{m}$, the trainable parameters of VQC are optimized by minimizing the following loss function:
\begin{equation}\label{eq:qsvdd_loss}
\mathcal{L}_{q}(\boldsymbol{\theta})=\frac{1}{m}\sum_{i=1}^{m}\|\varphi_{q}(\boldsymbol{x}_{i},\boldsymbol{\theta})-\boldsymbol{c}\|^{2},
\end{equation}
where $\boldsymbol{c}$ is a predefined center. Unlike Eq.~\ref{eq:dsvdd_loss}, the loss function for QSVDD does not include the regularization term, since the angle parameters of a quantum circuit naturally range from $0$ to $2\pi$. Similar to Eq.~\ref{eq:dsvdd_decision}, after training the feature mapping, whether a test data $\tilde{\boldsymbol{x}}\in\mathcal{X}$ is an oulier or not is determined by the decision function
\begin{equation}\label{eq:qsvdd_decision}
f(\tilde{\boldsymbol{x}})=\|\varphi_{q}(\tilde{\boldsymbol{x}};\boldsymbol{\theta}^{*})-\boldsymbol{c}\|^{2}-R^{2*}
\end{equation}
where $\boldsymbol{\theta}^{*}$ are the trained parameters and the radius $R^{2*}=\max_{i}\|\varphi_{q}(\boldsymbol{x}_{i};\boldsymbol{\theta}^{*})-\boldsymbol{c}\|^{2}$. Empirical studies have demonstrated the effectiveness of QSVDD, outperforming both classical~\cite{pmlr-v80-ruff18a} and quantum~\cite{Park_2023,Bravo_Prieto_2021} OCC methods.

\section{Model Description}
\begin{figure*}[t!] 
\centering
\includegraphics[,width=0.7\textwidth]{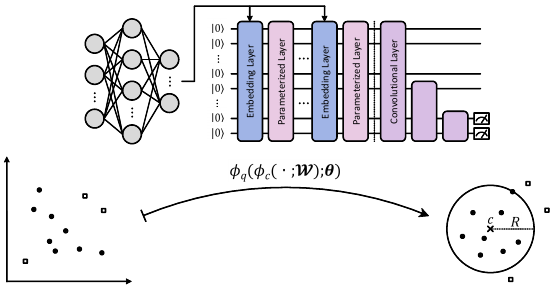}
\caption{
Overview of NQSVDD. Classical neural network layers extract features from complex input data. The output of classical layers is encoded into quantum state using ZZ feature embedding. Alternating repetition of the embedding layers and parameterized quantum layers makes the expressive power of the embedding richer. After the state preparation, variational quantum circuit and a set of Pauli measurements map the features into the latent space. The combination of classical layers, denoted as $\phi_{c}(\cdot;\boldsymbol{\mathcal{W}})$, and quantum layers, denoted as $\phi_{q}(\cdot;\boldsymbol{\theta})$, construct the overall feature mapping for OCC, trained to find the optimal latent space and hypersphere.
}
\label{fig:overview}
\end{figure*}

\subsection{Overview}

Although numerous QML-based OCC approaches have been proposed, some of these methods prioritize compressing data representations, using quantum autoencoder structures, over effective anomaly detection~\cite{Sakhnenko_2022,Ngairangbam_2022,Park_2023}. Instead of data compression, QSVDD, inspired from DSVDD, uses the objective for compact hypersphere encompassing the target data. However, exclusive reliance on quantum subroutines makes QSVDD ill-suited for processing modern complex datasets on current quantum hardware, from the standpoint of scalability and practicality.

Neural quantum support vector data description (NQSVDD) provides a practical approach by combining the representational power of quantum circuits with the robustness and scalability of classical neural network~\cite{Cerezo_2022}. Figure~\ref{fig:overview} provides a schematic overview of the NQSVDD architecture. The key components of the model are described in detail in the following subsections.

\subsection{Hybrid approach}

Under the constraints of current quantum hardware, the classical-quantum hybrid approach offers several advantages. Shallow Quantum neural networks (QNNs) learn quantum-correlated distributions, enhancing the representation capacity of classical models. At the same time, handling feature learning with classical neural networks alleviates the need for a deep, highly entangling ansatz, and many qubits~\cite{Cerezo_2022}. This makes the hybridization especially suitable for NISQ devices, where quantum resources are limited.

There are multiple ways to design hybrid models that combine quantum circuits with classical models, and the integration structure should be chosen with respect to the specific task and the nature of the data. In binary classification, the classical-to-quantum hybrid structure is proven to improve training accuracy, generalization, and robustness to noise~\cite{hur_NQE, hur_margin}. Extending this idea to OCC, NQSVDD adopts a classical-to-quantum hybrid structure for processing high-dimensional classical data. Specifically, a classical neural network initially compresses or preprocesses the input data, so that we can reduce the workload of QNN. The selection of the classical architecture is guided by inductive bias aligned with the structure of the data, e.g. spatial locality of convolutional neural networks(CNNs) for processing image data, and temporal causality of recurrent neural networks(RNNs) for sequential data. The output of classical neural network is then passed to the quantum circuit via some data embedding method. The specific structure of classical layers and quantum layers are designed considering the complexity of input data, the quantum hardware limitation, and the dimension of latent space.

\subsection{Data embedding}

Quantum data embedding is essential for processing classical data with quantum computer. Basis encoding, amplitude encoding, and angle encoding are typical methods of quantum embedding. Computational basis encoding maps classical bits directly to quantum basis states. It is simple and efficient, but not suitable for continuous data and not expressive enough. Amplitude encoding, which encodes a normalized data into the amplitudes of a quantum state, is compact and expressive, but the state preparation is inefficient, requiring complex subroutines. Angle encoding uses the data directly as rotation angles in single-qubit gates. It has shallow-depth circuits, but it is linear in the data. 

Our method is compatible with any quantum data embedding, but we adopt the ZZ-feature embedding~\cite{Havlicek_2019} for demonstration purposes, defined as follows:
\begin{equation}\label{eq:zzembedding}
    \begin{aligned}
        |\psi(\boldsymbol{x})\rangle &= U_{\text{ZZ}}(\boldsymbol{x})|0\rangle^{\otimes n}\\
        &:=\left\lbrack \left(\prod_{(j,k) \in \mathcal{P}} e^{i x_{jk} Z_j Z_k}\right)
        \left(\prod_{j=0}^{n-1} e^{i x_j Z_j}\right)
        \left(\bigotimes_{j=0}^{n-1} H_j\right)\right\rbrack^{l}
        |0\rangle^{\otimes n},
    \end{aligned}
\end{equation}
where \(\mathcal{P}\) denotes the set of qubit pairs used for entangling interactions. Here, assuming an even $l$ for simplicity, the $l$-layer ZZ-feature embedding can be interpreted as a first-order Lie-Trotter simulation of Hamiltonian $H_{\text{XZ}}(\boldsymbol{x})$~\cite{liu_2025}:
\begin{equation}\label{eq:zzembeddingtrotterization}
    \begin{aligned}
        &U_{\text{ZZ}}(\boldsymbol{x})\approx \left[e^{i(l/2)H_{\text{XZ}}(\boldsymbol{x})}\right],\\
        &\begin{aligned}
            \text{where}\quad H_{\text{XZ}}(\boldsymbol{x})&=\sum_{j=0}^{n-1}x_{j}(X_{j}+Z_{j})+\sum_{(j,k)\in\mathcal{P}}x_{jk}(X_{j}X_{k}+Z_{j}Z_{k}).
        \end{aligned}
    \end{aligned}
\end{equation}
In this Hamiltonian view, the input features are directly mapped to the local-field coefficients $x_{j}$ and the interaction-strength coefficients $x_{jk}$. For an \(n\)-qubit system, an $l$-layer ZZ-feature embedding can encode up to \(l(n+n(n-1)/2)=ln(n+1)/2\) features, considering all possible interaction pairs. The single-qubit gates $e^{i x_j Z_j}=\cos(x_{j})I+i\sin(x_{j})Z_{j}$ and two-qubit gates $e^{i x_{jk} Z_j Z_k}=\cos(x_{jk})I+i\sin(x_{jk})Z_{j}Z_{k}$ contribute nonlinearity of the embedding. This nonlinear dependence on the input data and the entanglement generated by the two-qubit interactions make the ZZ-feature embedding sufficiently expressive to capture complex data patterns. Furthermore, ZZ-feature embedding is related to the families of the instantaneous quantum polynomial (IQP) embedding, which can create highly entangled quantum states known to be classically hard to simulate~\cite{shepherd_2009, bremner_2011, bremner_2016}. By mapping classical data into an exponentially large Hilbert space through ZZ-feature embedding, NQSVDD can capture intricate feature representations that are computationally infeasible to construct classically, offering the potential to outperform conventional algorithms.

Additionally, NQSVDD alternates between ZZ-feature embedding layers \(U_{\text{ZZ}}(\boldsymbol{x})\) and parameterized unitary layers \(U(\theta_i)\).
\begin{equation}\label{eq:reuploading}
    |\psi(\boldsymbol{x})\rangle=[U(\theta_{t})\cdot U_{\text{ZZ}}(\boldsymbol{x})\cdots U(\theta_{1})\cdot U_{\text{ZZ}}(\boldsymbol{x})]|0\rangle^{\otimes n}.
\end{equation}
Repeatedly encoding classical input data into a quantum state, inserting trainable unitary operations between, further enhances expressiveness while maintaining shallow depth~\cite{PerezSalinas2020datareuploading}.

\subsection{Variational quantum circuit}

After preparing quantum state, a variational quantum circuit (VQC) $U(\boldsymbol{\theta})$ with parameters $\boldsymbol{\theta}=\{\theta_{1},...,\theta_{T}\}$ learns meaningful representations of embedded features in the exponentially large Hilbert space. The architecture of VQC strongly influences trainability, inductive bias, and final performance~\cite{McClean_2018, Sim_2019, Cerezo_2021_barren_plateaus, Wang_2021}. A well-designed architecture can preserve useful gradients, well representing the solution space, leading to a good trainability and generalization performance. On the other hand, a careless design can induce barren plateaus and overfit the noise. 

The ultimate goal of Deep SVDD is to learn a mapping that embeds inputs inside a compact hypersphere. Practically, the hypersphere is defined in a lower-dimensional latent space to improve the generalization performance of OCC, implicitly removing redundant variability while retaining distinctive features of the target class distribution relative to the others. This objective naturally leads to a hierarchical architecture of VQC with local measurement, which preserves important information during progressive compression~\cite{Grant_2018}. Quantum convolutional neural network (QCNN), a representative hierarchical VQC, consist of two layer types: convolutional layers that apply local unitary gates for data processing, and pooling layers that reduce the number of active qubits by measuring or discarding certain qubits. This repetitive reduction in system size of the hierarchical structure of QCNN efficiently compresses an $n$-qubit input to a constant number of qubits in $O(\log n)$ depth. As a result of the compression, we can choose low-weight ($k$-local) Pauli observables with only a small number of non-identity Pauli factors, enabling efficient expectation estimation via classical shadow tomography~\cite{Huang_2020}. Trainability and good generalization also motivate the QCNN architecture. The hierarchical and locally connected architecture of QCNN prevents the issue of barren plateaus, guaranteeing trainability and efficiency~\cite{pesah2020absence}. Moreover, a theoretical analysis and empirical experiments show strong generalization capabilities and excellent binary classification performance~\cite{Banchi_2021,hur_QCNN,kim2023classical}.

\begin{figure}[!t]
\centering
\includegraphics[width=.5\textwidth]{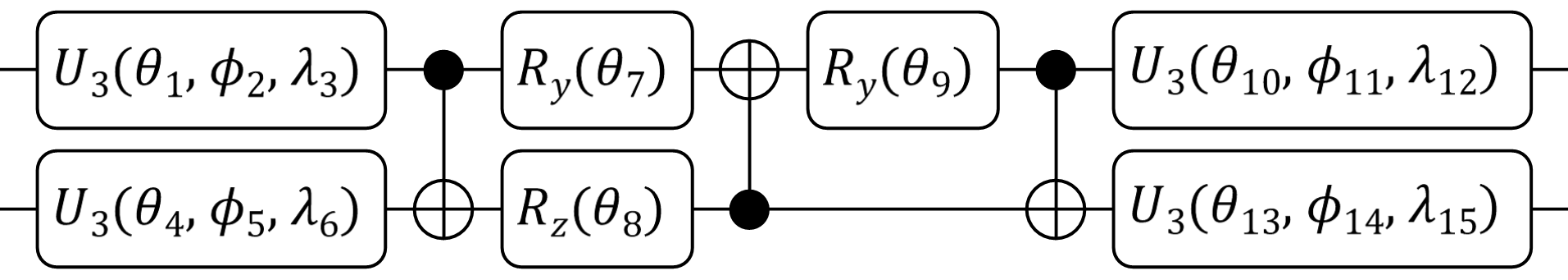}
\caption{The parameterized two-qubit gate that can generate an arbitrary two-qubit unitary transformation in $SU(4)$, requiring $15$ real parameters. \(R_{i}(\theta)\) is a rotation gate of \(i\) axis of Bloch sphere, and \(U_{3}(\theta, \phi, \lambda)=R_{z}(\phi)R_{x}(-\pi/2)R_{z}(\theta)R_{x}(\pi/2)R_{z}(\lambda)\). This two-qubit ansatz is applied for parameterized unitary layers in data embedding, and convolutional layers in QCNN.}
\label{fig:su4}
\end{figure}

For the parameterized two-qubit gates, applied for trainable unitary layers of data embedding and convolutional layers of QCNN, we adopt an \(SU(4)\) ansatz, which requires 15 real parameters and can generate an arbitrary two-qubit unitary operation (see Fig.~\ref{fig:su4}). All two-qubit gates within each layer share the same parameters, which improves generalization performance and reduces the number of trainable parameters~\cite{liu2017survey}. Since the depth of QCNN layers scales logarithmically to the number of qubits, a QCNN can be characterized by $O(\log n)$ number of parameters. 

The final quantum state $|\psi_{f}\rangle=U(\boldsymbol{\theta})|\psi(x)\rangle$ is projected into a latent space \(\mathcal{F}\) through local measurements, and the expectation values of selected observables define the output features. Specifically, the final quantum state, which exists on \(n_f\) active qubits, can be fully characterized by a density matrix. This density matrix requires \(4^{n_f} - 1\) real parameters to be completely described. These features are extracted by measuring the expectation values of a selected subset \(P \subseteq \mathcal{P}_{n_f}\) of Pauli observables, where \(\mathcal{P}_{n_f} = \{I, \sigma_x, \sigma_y, \sigma_z\}^{\otimes n_f}\setminus \{ I^{\otimes n_f} \}\). The \(i\)th output (latent) feature is computed as the expectation value \(\langle P_i \rangle\), where \(P_i\in P\subseteq\mathcal{P}_{n_{f}}\) is a chosen Pauli observable. 
\begin{equation}
\begin{aligned}
    \langle P_{i}\rangle&=\langle\psi_{f}|P_{i}|\psi_{f}\rangle=\text{Tr}(P_{i}|\psi_{f}\rangle\langle\psi_{f}|)\\
    &=\text{Tr}(P_{i}U(\boldsymbol{\theta})|\psi(x)\rangle\langle\psi(x))| U^{\dagger}(\boldsymbol{\theta}))=\text{Tr}(U^{\dagger}(\boldsymbol{\theta})P_{i}U(\boldsymbol{\theta})|\psi(x)\rangle\langle\psi(x))|).
\end{aligned}
\end{equation}
This indicates that optimizing a VQC can be interpreted as learning optimal parameterized measurements. Since the density matrix of the final state is fully characterized by $4^{n_{f}}-1$ real parameters, the maximum possible number of output features is $4^{n_{f}}-1$. In order to force the model to capture only the most important features of target data distribution rather than simply memorizing the noise pattern of each sample, the dimension of the latent feature space $d'$, determined by the number of elements in a set $P$ of selected Pauli observables, should satisfy $d'=|P|\leq\min\{d, 4^{n_{f}}-1\}$. Additionally, if the selected set $P$ consists of low-weight Pauli observables, classical shadow tomography can efficiently estimate their expectation values~\cite{Huang_2020}. Moreover, grouping qubit-wise commuting Pauli terms can further improve the efficiency of multiple Pauli measurements~\cite{Verteletskyi_2020}.

\subsection{Loss function}

The loss function for NQSVDD is a combination of Eq.~\ref{eq:dsvdd_loss} and Eq.~\ref{eq:qsvdd_loss}. Let \(\boldsymbol{z}=\phi_c(\boldsymbol{x}; \boldsymbol{\mathcal{W}})\) be the output of a classical NN with weights \(\boldsymbol{\mathcal{W}} = \{ \boldsymbol{W}^1, \ldots, \boldsymbol{W}^L \}\) across \(L\) layers, and  \(\phi_q(\boldsymbol{z}; \boldsymbol{\theta}) = g(U(\boldsymbol{\theta})|\psi(\boldsymbol{z})\rangle)\) be the output of the quantum model with parameters $\boldsymbol{\theta}=\{\theta_{1},...,\theta_{T}\}$, where \(|\psi(\boldsymbol{z})\rangle\) is the encoded quantum state and \(g:\mathbb{C}^{2^{n}}\rightarrow\mathcal{F}\) is a function that maps the quantum state to a latent space. The overall feature mapping is expressed
\(
\phi(\boldsymbol{x}; \boldsymbol{\mathcal{W}}, \boldsymbol{\theta}) = \phi_q\left( \phi_c(\boldsymbol{x}; \boldsymbol{\mathcal{W}}); \boldsymbol{\theta} \right)
\). 
Given examples \(\{ \boldsymbol{x}_i \}_{i=1}^m\), the model is trained to map inputs close to a reference center \(\boldsymbol{c} \in \mathcal{F}\) by minimizing the following objective:
\begin{equation} \label{eq:hqsvdd}
\mathcal{L}(\boldsymbol{\mathcal{W}}, \boldsymbol{\theta}) =
\frac{1}{m} \sum_{i=1}^m \left\| \phi(\boldsymbol{x}_i; \boldsymbol{\mathcal{W}}, \boldsymbol{\theta}) - \boldsymbol{c} \right\|^2
+ \frac{\lambda}{2} \sum_{\ell=1}^L \left\| \boldsymbol{W}^\ell \right\|_F^2.
\end{equation}
The first term encourages compactness around the center \(\boldsymbol{c}\) by minimizing the average of the squared Euclidean distance between the embedded output and the center over $m$ samples. The second term is a Frobenius norm regularization with hyperparameter \(\lambda > 0\). Since quantum parameters \(\boldsymbol{\theta}\in[0,2\pi]\) are inherently bounded, the regularization is only applied to classical parameters of $L$ layers classical NN.

\subsection{Gradient-based optimization of hybrid model}

In the step of gradient-based optimization, the parameters of classical and quantum neural networks are jointly updated. The analytic gradients of expectation values in parameterized quantum circuits with respect to quantum parameters are computed using parameter-shift rule~\cite{Schuld_2019}. Let $|\psi_{\boldsymbol{\theta}}\rangle=U(\boldsymbol{\theta})|\psi\rangle$ be a parameterized quantum state, where $U(\boldsymbol{\theta})=e^{-i\boldsymbol{\theta}G}$ with $G^{2}=I$ and $\boldsymbol{\theta}=\{\theta_{1},...,\theta_{T}\}$, and $|\psi\rangle$ is an initial quantum state. Then, for a Hermitian observable $\hat{O}$, the expectation value is $\langle\hat{O}\rangle_{\boldsymbol{\theta}}=\langle\psi_{\boldsymbol{\theta}}|\hat{O}|\psi_{\boldsymbol{\theta}}\rangle$. The parameter-shift rule gives the exact gradient of $\langle\hat{O}\rangle_{\boldsymbol{\theta}}$ with respect to $\theta_{i}$:
\begin{equation}\label{eq:parameter_shift_rule}
    \frac{\partial\langle\hat{O}\rangle_{\boldsymbol{\theta}}}{\partial\theta_{i}}=\frac{1}{2}\left\lbrack\langle\hat{O}\rangle_{\boldsymbol{\theta}^{+}_{i}}-\langle\hat{O}\rangle_{\boldsymbol{\theta}_{i}^{-}}\right\rbrack,
\end{equation}
where $\boldsymbol{\theta}_{i}^{\pm}=(\theta_{1},...,\theta_{i}\pm\frac{\pi}{2},...,\theta_{T})$ and $\langle\hat{O}\rangle_{\boldsymbol{\theta}^{\pm}_{i}}=\langle\psi_{\boldsymbol{\theta}^{\pm}_{i}}|\hat{O}|\psi_{\boldsymbol{\theta}^{\pm}_{i}}\rangle$. In case of classical neural networks, the gradients are computed by recursively applying the chain rule during backpropagation to propagate the loss derivative from the output layer back through each layer. Computing gradients of a hybrid model can be done by combining the parameter-shift rule for quantum parameters and backpropagation for classical parameters via the chain rule. Consider a hybrid model of classical-to-quantum structure, where the output of classical neural network $\boldsymbol{z}$, with classical parameters $\boldsymbol{w}$ and quantum parameters $\boldsymbol{\theta}$. For some loss $\mathcal{L}$, the gradients with respect to $\theta_{i}$ and $w_{j}$ are computed by chain rule, expressed as:
\begin{equation}\label{ep:hybrid_gradient}
    \begin{aligned}
        &\frac{\partial \mathcal{L}}{\partial\theta_{i}}=\frac{\partial \mathcal{L}}{\partial\langle\hat{O}\rangle_{\boldsymbol{\theta}}}\cdot\frac{\partial\langle\hat{O}\rangle_{\boldsymbol{\theta}}}{\partial\theta_{i}},\\
        &\frac{\partial \mathcal{L}}{\partial w_{j}}=\frac{\partial \mathcal{L}}{\partial\langle\hat{O}\rangle_{\boldsymbol{\theta}}}\cdot\sum_{k}\left(\frac{\partial\langle\hat{O}\rangle_{\boldsymbol{\theta}}}{\partial z_{k}}\cdot\frac{\partial z_{k}}{\partial w_{j}}\right),
    \end{aligned}
\end{equation}
where $\partial\langle\hat{O}\rangle_{\boldsymbol{\theta}}/\partial\theta_{i}$ and $\partial\langle\hat{O}\rangle_{\boldsymbol{\theta}}/\partial z_{k}$ are obtained by applying the parameter-shift rule. By computing $\partial\mathcal{L}/\partial\theta_{i}$ and $\partial\mathcal{L}/\partial w_{j}$, it is possible to jointly optimize all classical and quantum parameters.

\section{Numerical experiments}
\noindent Experiments are conducted on the MNIST~\cite{726791}, Fashion-MNIST~\cite{xiao2017fashionmnist}, Credit Card Transaction~\cite{kagglecredit} dataset, and Network Intrusion (CIC-IDS2017)~\cite{kagglenetwork} dataset. We first compare the OCC performance of NQSVDD with that of baseline approaches on each dataset under ideal, noiseless setting. We additionally evaluate its effectiveness of OCC tasks on a noisy environment. Detailed descriptions of the datasets and experimental procedures are provided in the following sections.

\subsection{Experiments on MNIST and Fashion-MNIST}

MNIST and Fashion-MNIST dataset are $28\times28$ pixels grayscale images of handwritten digits and clothing items, respectively, both having $10$ different classes. Each class is treated target, resulting in \(10\) independent OCC tasks, each with a different target class. NQSVDD is trained on $1000$ target samples and evaluated using the AUC score with $100$ target and $90$ outlier test samples ($10$ samples from each remaining class). All images are pre-processed with normalization and rescaling to $[0,1]$ via min-max scaling.

The classical layers of NQSVDD has two alternating convolutional layers and pooling layers. The degree of reduction in pooling layer is determined by considering the dimension of features encoded to quantum state via ZZ-feature embedding: $36$-dimensional features in eight-qubit state. Additionally, NQSVDD maps the flattened output of classical layers by alternating ZZ-feature embedding layers and parameterized unitary layers three times. The results of increasing the number of embedding layers are provided in Fig.~\ref{fig:byemb} of Appendix~\ref{appendix}. After the state preparation, two parameterized layers on eight qubits and two layers on four qubits compose QCNN, followed by a set of Pauli measurements. As mentioned above, all $SU(4)$ ansatz within each quantum layer share the same parameters. Also, the dimension of the latent space should be selected carefully since it directly affects the performance of the model, as shown in Fig.~\ref{fig:bylatent_mnist} and ~\ref{fig:bylatent_fashion} of Appendix~\ref{appendix}. In order to compete against the conventional DSVDD that sets the latent dimension to $32$, we selected a set of $32$ Pauli observables in $\mathcal{P}_{4}=\{I,\sigma_{x},\sigma_{y},\sigma_{z}\}^{\otimes4}\setminus \{ I^{\otimes 4} \}$.

\begin{figure}
    \centering
    \subfloat[MNIST\label{fig:auc_mnist}]{
    \includegraphics[width=0.45\textwidth]{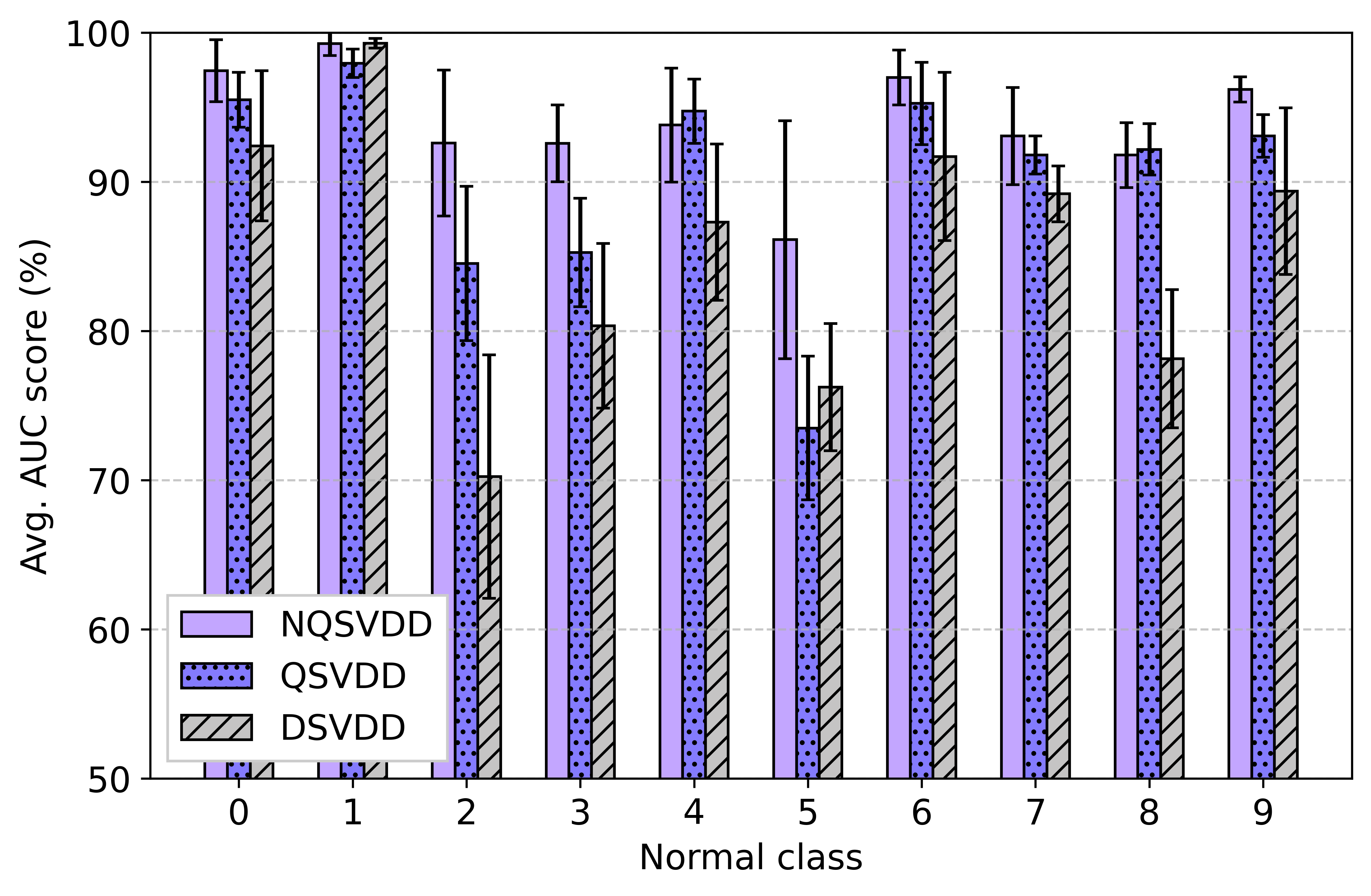}
    }
    \hfill
    \subfloat[Fashion-MNIST\label{fig:auc_fashion}]{
    \includegraphics[width=0.45\textwidth]{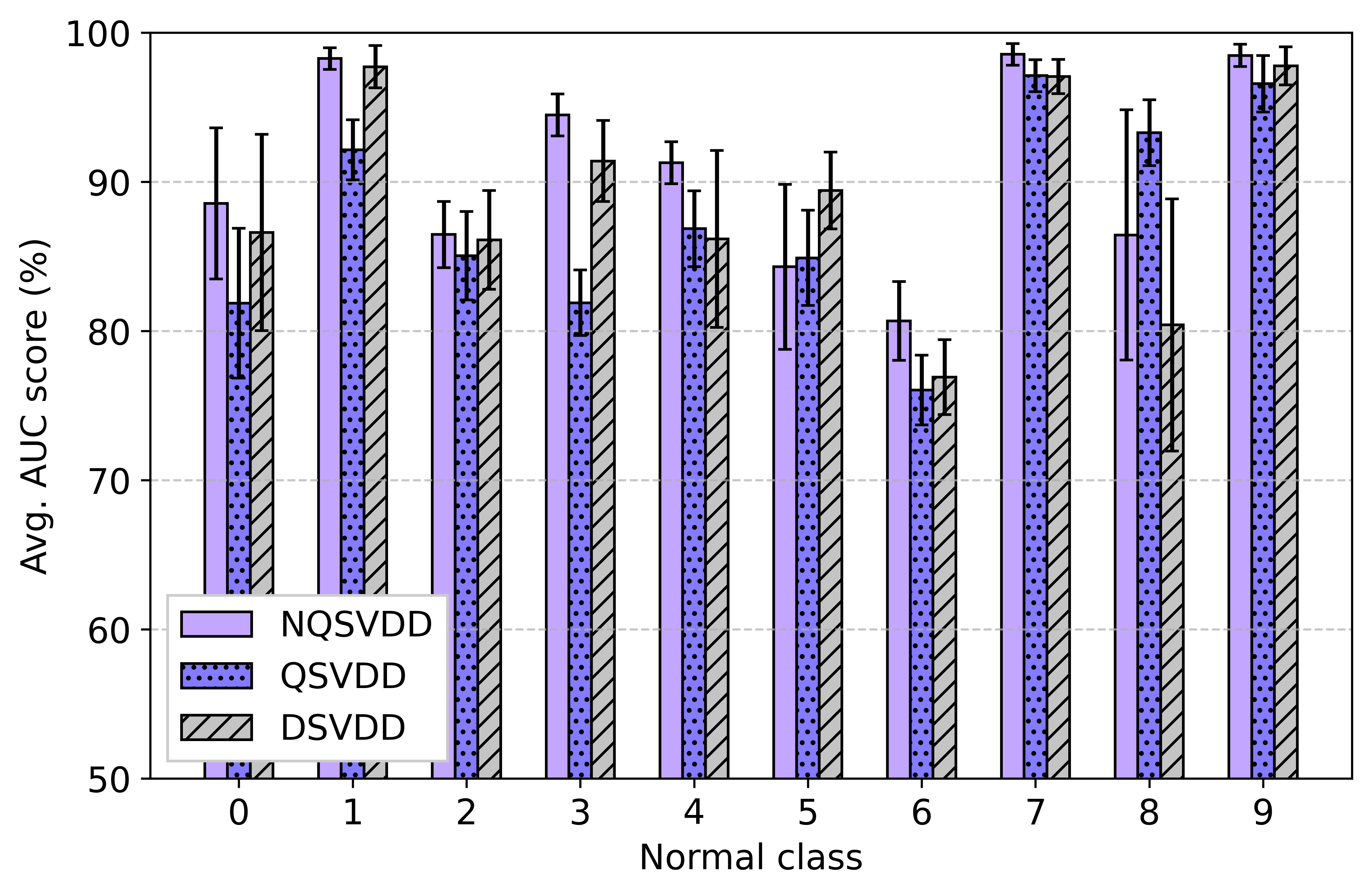}
    }
    \caption{
    Avg. AUC (\%) and std. (error bar) on (a) MNIST and (b) Fashion-MNIST datasets. NQSVDD, QSVDD and DSVDD have \(1105\), \(75\) and \(2152\) number of parameters, respectively. Each bar represents the average score of \(10\) repetitions.
    }
\end{figure}

We compare the proposed NQSVDD with QSVDD and the classical DSVDD. The two QSVDD variants, QSVDD with amplitude encoding and QSVDD with ZZ feature embedding, share an identical variational circuit structure of \(75\) trainable parameters, while input data is resized and flattened according to the type of encoding. For DSVDD, we consider the original architecture from ~\cite{pmlr-v80-ruff18a} and the reduced one is modified so that its convolutional and pooling layers match those of NQSVDD. These two DSVDD models have \(7272\) and \(2152\) parameters, respectively, which is roughly \(7\) and \(2\) times of the number of parameters in NQSVDD, \(1105\). NQSVDD and QSVDD are implemented on \(8\) input qubits using PennyLane~\cite{bergholm2020pennylane}. The parameter sets $\boldsymbol{\mathcal{W}}$ and $\boldsymbol{\theta}$ are updated simultaneously via Adam optimizer~\cite{kingma2017adammethodstochasticoptimization} with learning rate $0.05$. We also use cosine annealing with warm restarts~\cite{loshchilov2017sgdrstochasticgradientdescent}, which decays the learning rate to a minimum of \(0.005\) and periodically resets to improve convergence and escape local minima. We trained $1500$ steps using a batch size of $32$. Figures~\ref{fig:auc_mnist}, ~\ref{fig:auc_fashion}, and Tab.~\ref{tab:mnist} show that NQSVDD outperforms both QSVDD and DSVDD across most tasks. Notably, NQSVDD uses significantly fewer parameters than DSVDD, yet consistently achieves performance on par with or superior to DSVDD.

\begin{table}[h]
    \begin{center}
        \caption{Average AUC scores in \% with StdDevs with 10 random seeds per method and one-class experiment on MNIST and Fashion-MNIST dataset. Boldface indicates the best performing result within each class.}
        \label{tab:mnist}
        \setlength{\tabcolsep}{10pt}
        \renewcommand{\arraystretch}{1.1}
        \begin{tabular}{lccccc}
            \toprule
            & & & Method & &\\
            \cline{2-6}
            Target Class & NQSVDD  & QSVDD(amp) & QSVDD(zz) & DSVDD & DSVDD(reduced) \\
            \hline
            0 & $\boldsymbol{97.45}\pm2.07$ & $95.50\pm1.84$ & $88.70\pm3.49$ & $94.49\pm3.80$ & $92.42\pm5.03$ \\
            1 & $99.27\pm0.79$ & $97.96\pm0.95$ & $97.36\pm1.36$ & $\boldsymbol{99.54}\pm0.40$ & $99.29\pm0.32$ \\
            2 & $\boldsymbol{92.60}\pm4.89$ &  $84.53\pm5.17$ & $82.70\pm2.87$ & $77.48\pm10.84$ & $70.25\pm8.16$ \\
            3 & $\boldsymbol{92.59}\pm2.58$ &  $85.27\pm3.64$ & $83.71\pm4.48$ & $80.28\pm8.60$ & $80.35\pm5.52$ \\
            4 & $93.81\pm3.83$ &  $\boldsymbol{94.74}\pm2.15$ & $88.32\pm4.01$ & $90.54\pm2.88$ & $87.29\pm5.24$ \\
            5 & $\boldsymbol{86.13}\pm7.98$ &  $73.50\pm4.82$ & $78.51\pm4.71$ & $76.42\pm10.33$ & $76.23\pm4.27$ \\
            6 & $\boldsymbol{96.99}\pm1.84$ &  $95.26\pm2.76$ & $85.27\pm2.36$ & $94.90\pm2.13$ & $91.70\pm5.64$ \\
            7 & $\boldsymbol{93.08}\pm3.26$ &  $91.80\pm1.27$ & $83.63\pm4.18$ & $91.88\pm3.42$ & $78.15\pm4.63$ \\
            8 & $91.80\pm2.18$ &  $\boldsymbol{92.18}\pm1.72$ & $82.61\pm3.41$ & $82.64\pm3.65$ & $78.15\pm4.63$ \\
            9 & $\boldsymbol{96.20}\pm0.85$ &  $93.08\pm1.43$ & $87.38\pm4.95$ & $92.45\pm4.19$ & $89.39\pm5.58$ \\
            \hline
            Total Avg. & $\boldsymbol{93.99}\pm4.98$ &  $90.38\pm7.53$ & $85.82\pm6.02$ & $88.06\pm9.69$ & $85.43\pm9.73$\\
            \midrule
            T-shirt/top & $\boldsymbol{88.56}\pm5.07$ &  $81.87\pm5.01$ & $85.24\pm3.47$ & $87.60\pm6.53$ & $86.60\pm6.58$ \\
            Trouser & $\boldsymbol{98.27}\pm0.72$ &  $92.15\pm2.00$ & $96.14\pm1.68$ & $97.38\pm1.53$ & $97.72\pm1.41$ \\
            Pullover & $86.47\pm2.21$ & $85.05\pm2.96$ & $\boldsymbol{87.87}\pm1.28$ & $87.59\pm1.43$ & $86.11\pm3.31$ \\
            Dress & $\boldsymbol{94.48}\pm1.41$ &  $81.90\pm2.21$ & $89.90\pm3.65$ & $93.24\pm1.62$ & $91.40\pm2.72$ \\
            Coat & $\boldsymbol{91.29}\pm1.40$ &  $86.87\pm2.54$ & $83.91\pm1.74$ & $89.22\pm3.66$ & $86.18\pm5.93$ \\
            Sandal & $84.31\pm5.53$ &  $84.90\pm3.19$ & $\boldsymbol{91.01}\pm1.97$ & $88.42\pm5.67$ & $89.42\pm2.58$ \\
            Shirt & $\boldsymbol{80.68}\pm2.63$ &  $76.05\pm2.33$ & $80.52\pm3.49$ & $75.50\pm4.48$ & $76.91\pm2.51$ \\
            Sneaker & $\boldsymbol{98.55}\pm0.73$ &  $97.12\pm1.07$ & $97.42\pm1.06$ & $97.30\pm1.81$ & $97.06\pm1.15$ \\
            Bag & $86.44\pm8.39$ &  $\boldsymbol{93.30}\pm2.20$ & $78.44\pm5.22$ & $85.08\pm3.98$ & $80.41\pm8.45$ \\
            Ankle boot & $\boldsymbol{98.48}\pm0.75$ &  $96.58\pm1.89$ & $92.94\pm2.31$ & $97.03\pm2.47$ & $97.77\pm1.28$\\
            \hline
            Total Avg. & $\boldsymbol{90.75}\pm7.15$ &  $87.58\pm7.14$ & $88.34\pm6.64$ & $89.83\pm7.41$ & $88.96\pm8.00$\\
            \bottomrule
        \end{tabular}
    \end{center}
\end{table}

\subsection{Experiments on Credit Card Transaction} 

Credit Card Transaction dataset contains transactions made by European cardholders over two days in September 2013, having $492$ frauds out of $284807$ transactions~\cite{kagglecredit}. Due to confidentiality issues, original features are not provided. Instead, numerical features of $28$ dimension are given as the result of a PCA transformation. Similar to MNIST and Fashion-MNIST, randomly selected $1000$ target samples are used for training, and the AUC score is computed using $492$ target samples and $492$ outlier samples. 

Given that the Credit Card Transaction data samples have lower complexity than the previous image dataset, we scaled down the model architectures accordingly. In case of NQSVDD, classical backbone of the model consists of two blocks that alternate 1D convolution and pooling layer, transforming the $28$-dimensional input into a compact $21$-dimensional feature vector. By repeating three blocks of ZZ-feature embedding and parameterized layers, this feature vector is encoded into a six-qubit register, which is hierarchically compressed by QCNN into two active qubits. Local Pauli measurements finally yield an eight-dimensional latent representation. QSVDD and DSVDD are likewise scaled to produce an eight-dimensional latent vector, with DSVDD sharing the same convolutional backbone as NQSVDD. The total counts of trainable parameters in NQSVDD, QSVDD, DSVDD are $210$, $75$, and $288$, respectively. Other settings of experiment are identical to those on MNIST and Fashion-MNIST. Table~\ref{tab:creditcard} shows that NQSVDD outperforms both QSVDD and DSVDD.

\begin{table}[h]
    \begin{center}
        \caption{Average AUC scores in \% with StdDevs with 20 random seeds per method on Credit Card Transaction dataset.}
        \label{tab:creditcard}
        \setlength{\tabcolsep}{10pt}
        \renewcommand{\arraystretch}{1.2}
        \centering
        \begin{tabular}{ccc}
            \toprule
            Method & Params & Avg. AUC \\
            \hline
            NQSVDD & $210$ & $\boldsymbol{94.67} \pm 0.84$ \\
            QSVDD(amp) & $75$ & $87.98 \pm 7.59$ \\
            DSVDD & $288$ & $92.30 \pm 2.24$ \\
            \bottomrule
        \end{tabular}
    \end{center}
\end{table}

\subsection{Experiments on Network Intrusion}

CIC-IDS2017 dataset is a network intrusion benchmark released by the Canadian Institute for Cybersecurity (CIC) in 2017 for the evaluation of the Intrusion Detection System (IDS)~\cite{kagglenetwork}. The dataset captures several days of mixed benign and malicious traffic, extracting $78$ network flow features. In this experiment, we selected the dataset of Thursday, July 6, 2017, morning, which contains four classes: ``Benign", ``Web Attack - Brute Force", ``Web Attack - Sql Injection", ``Web Attack - XSS", each with $168166$, $1507$, $652$, and $21$ samples, respectively. Because our objective is to detect whether a web attack occurred, the three web attack classes were merged into a single ``Web Attack" label. For training, $1000$ ``Benign" samples were randomly sampled, and the AUC score is computed using a balanced test set of $2180$ ``Benign" samples and $2180$ ``Web Attack" samples. 

The models are downsized analogously to those used in the Credit Card Transaction dataset experime-ts. The cal lassical backbone of NQSVDD extracts $21$-dimensional feature vector from $78$-dimensional input. Three repetitions of ZZ-feature embeddings and parameterized layers encode this vector into six-qubit system, followed by QCNN with the final three active qubits, producing $16$-dimensional latent representation. NQSVDD, QSVDD and DSVDD contain $225$, $75$, and $456$ trainable parameters, respectively. As shown in Table~\ref{tab:network}, NQSVDD outperforms both QSVDD and DSVDD.

\begin{table}[h]
    \begin{center}
        \caption{Average AUC scores in \% with StdDevs with 20 random seeds per method on Network Intrusion dataset.}
        \label{tab:network}
        \setlength{\tabcolsep}{10pt}
        \renewcommand{\arraystretch}{1.2}
        \begin{tabular}{ccc}
            \toprule
            Method & Params & Avg. AUC \\
            \hline
            NQSVDD & $225$ & $\boldsymbol{75.32} \pm 10.58$ \\
            QSVDD(amp) & $75$ & $56.04 \pm 14.35$ \\
            DSVDD  & $456$ & $59.45 \pm 23.46$ \\
            \bottomrule
        \end{tabular}
    \end{center}
\end{table}

\subsection{Noisy Simulation}

\begin{table}[t]
    \begin{center}
        \setlength{\tabcolsep}{10pt}
        \renewcommand{\arraystretch}{1.2}
        \caption{Median error rates and gate lengths for the IBM quantum device ibm\_kingston used in the noisy simulation. $T_{1}$ and $T_{2}$ represents energy relaxation time and dephasing time, respectively.}
        \label{tab:noise}
        \begin{tabular}{cc}
            \toprule
            Backend Params &  \\
            \hline
            2-qubit Depolarizing & $0.00332$ \\
            1-qubit Gate Length (ns) & $32$ \\
            2-qubit Gate Length (ns) & $68$ \\
            $T_{1}$ ($\mu$s) & $183.29$ \\
            $T_{2}$ ($\mu$s) & $141.73$ \\
            \bottomrule
        \end{tabular}
    \end{center}
\end{table}

\begin{table*}[t]
    \begin{center}
        \caption{Average AUC scores in \% with StdDevs with five random seeds per method on MNIST, Fashion MNIST, Credit Card Transaction, Network Intrusion datasets under both noiseless and noisy conditions. The noisy NQSVDD is trained for $500$ iterations with a batch size of four. The noiseless NQSVDD and the classical DSVDD adopt the same suboptimal configuration.}
        \label{tab:noise_auc}
        \setlength{\tabcolsep}{10pt}
        \renewcommand{\arraystretch}{1.2}
        \begin{tabular}{ccccc}
            \toprule
            Dataset & Method & Params & Condition & Avg. AUC \\
            \hline
            \multirow{3}{*}{\makecell{MNIST\\(target: 0)}}
             & NQSVDD   & $1090$    & Noiseless & $97.92\pm1.33$    \\
             &          &           & Noise     & $94.00\pm2.37$    \\
             \cline{2-4}
             & DSVDD    & $2152$    & -         & $92.11\pm5.11$    \\
            \midrule
            \multirow{3}{*}{\makecell{Fashion-MNIST\\(target: T-shirt)}}
             & NQSVDD   & $1090$    & Noiseless & $90.14\pm2.75$    \\
             &          &           & Noise     & $85.84\pm1.83$    \\
             \cline{2-4}
             & DSVDD    & $2152$    & -         & $81.12\pm6.69$    \\
             \midrule
            \multirow{3}{*}{\makecell{Credit Card\\Transaction}}
             & NQSVDD   & $195$     & Noiseless & $94.19\pm1.07$    \\
             &          &           & Noise     & $93.00\pm2.17$    \\
             \cline{2-4}
             & DSVDD    & $456$     & -         & $89.83\pm2.50$    \\
             \midrule
            \multirow{3}{*}{\makecell{Network\\Intrusion}}
             & NQSVDD   & $195$     & Noiseless & $70.67\pm14.17$   \\
             &          &           & Noise     & $64.50\pm21.38$   \\
             \cline{2-4}
             & DSVDD    & $456$     & -         & $56.10\pm16.02$   \\
            \bottomrule
        \end{tabular}
    \end{center}
\end{table*}

In order to verify the effectiveness of our method on NISQ devices, we performed OCC experiments under a noisy environment, using noise parameters from the IBM quantum device ibm\_kingston. The noise model contains two-qubit depolarizing noise and thermal relaxation noise. The two-qubit depolarizing channel with error probability \(p\in[0,1]\) applies one of the fifteen non-identity two-qubit Pauli errors uniformly at random with total probability \(p\). The thermal relaxation error models the amplitude and phase damping with the energy relaxation time $T_{1}$, the dephasing time $T_{2}$, and the gate duration (length) $t$. Specific values for two-qubit depolarizing noise and thermal relaxation noise are provided in Table~\ref{tab:noise}. Single-qubit depolarizing errors and readout errors were excluded as their impact is substantially smaller than that of two-qubit errors and effective measurement error mitigation techniques are available~\cite{Bravyi_2021, Nation_2021, Kwon_2021, van_den_Berg_2022, Kim_2022, Lee_2023}. 

We evaluated NQSVDD on noisy simulation and compared its OCC performance with an ideal, noiseless simulation of NQSVDD and the classical DSVDD baseline. To prevent excessive circuit depth in NQSVDD, we restricted the number of embedding layers to two. Each method was trained and evaluated on four OCC tasks, using the following target classes: ``0" class from MNIST, ``T-shirt" class from Fashion-MNIST, ``Non-fraud" class from Credit Card Transaction dataset, and ``Benign" class from Network Intrusion dataset. Training was conducted for 500 iterations with a batch size of four, while retaining the rest of the settings from the preceding ideal simulations. To align with the noisy-simulation setting of NQSVDD, we conducted the experiments using suboptimal hyperparameters, specifically the batch size and the number of iterations, for noiseless NQSVDD and DSVDD. For each dataset, we report the average AUC and the sample standard deviation over five different random seeds per method.

Detailed results for noisy setting are provided in Table~\ref{tab:noise_auc}. NQSVDD under hardware noise consistently underperforms its noiseless version, indicating the impact of device noise on quantum model's effectiveness. Nevertheless, even in the noisy regime, NQSVDD consistently exceeds the classical DSVDD baseline across all datasets while using roughly half the number of parameters. These results suggest that, in the NISQ era, quantum approaches to OCC can be competitively effective compared to established classical approaches.

\section{Conclusion}
In this study, we introduced Neural Quantum Support Vector Data Description (NQSVDD), a classical–quantum hybrid learning framework for one-class classification (OCC). The proposed framework integrates classical neural networks with quantum data embedding and variational quantum circuits in a unified optimization scheme. Through this hybrid architecture, input data are first transformed by a classical neural network to optimize their quantum feature mapping into a high-dimensional Hilbert space, and processed through interleaved data re-uploading and trainable variational quantum circuits, yielding a compact latent representation defined by quantum measurements and optimized for the SVDD objective. This hybrid design enables expressive feature learning while maintaining low circuit depth and parameter efficiency, making it suitable for near-term quantum devices.

Evaluations on MNIST, Fashion MNIST, Credit Card Transaction, and Network Intrusion datasets demonstrated the effectiveness of NQSVDD. The proposed model consistently outperformed or at least matched both classical DSVDD and QSVDD baselines in terms of AUC performance, even with fewer trainable parameters. Furthermore, even under noisy quantum simulations, NQSVDD maintained robust OCC performance relative to the classical method, highlighting its resilience to quantum hardware noise and suitability for practical deployment on near-term quantum platforms.

Overall, these results indicate that NQSVDD effectively combines the expressive capabilities of quantum feature transformations with the stability and scalability of classical neural networks, providing a practical hybrid framework for OCC. Interesting future research directions include exploring gradient-free optimization techniques\cite{Bonet_Monroig_2023}, refining quantum embedding strategies for higher-dimensional data, and extending the proposed framework to semi-supervised and multi-class scenarios. Moreover, validating the model on large-scale quantum processors would provide valuable insights into its scalability and robustness in practical applications. Finally, since the initial classical neural networks can be designed to suit specific data domains, it would be an additional direction to investigate the effectiveness of using different classical models for different domains.

\section*{Acknowledgments}
\noindent This work was supported by a Korea Research Institute for Defense Technology Planning and Advancement grant funded by Defense Acquisition Program Administration (DAPA) (KRIT-CT-23-031). This work was also supported by the Yonsei University Research Grant of 2025, Institute of Information \& communications Technology Planning \& evaluation (IITP) grant funded by the Korea government (No. 2019-0-00003, Research and Development of Core technologies for Programming, Running, Implementing and Validating of Fault-Tolerant Quantum Computing System), the National Research Foundation of Korea (RS-2025-02309510), and the Ministry of Trade, Industry, and Energy (MOTIE), Korea, under the Industrial Innovation Infrastructure Development Project (Project No. RS-2024-00466693).

\newpage

\begin{thebibliography}{10}

\bibitem{Cerezo_2022}
M.~Cerezo, G.~Verdon, H.-Y. Huang, L.~Cincio, and P.~J. Coles, ``Challenges and opportunities in quantum machine learning,'' {\em Nat. Comput. Sci.}, vol.~2, pp.~567--576, 2022.

\bibitem{Schuld_2014}
M.~Schuld, I.~Sinayskiy, and F.~Petruccione, ``An introduction to quantum machine learning,'' {\em Contemporary Physics}, vol.~56, pp.~172--185, 2014.

\bibitem{Biamonte_2017}
J.~Biamonte, P.~Wittek, N.~Pancotti, P.~Rebentrost, N.~Wiebe, and S.~Lloyd, ``Quantum machine learning,'' {\em Nature}, vol.~549, pp.~195--202, 2017.

\bibitem{Rebentrost_2014}
P.~Rebentrost, M.~Mohseni, and S.~Lloyd, ``Quantum support vector machine for big data classification,'' {\em Phys. Rev. Lett.}, vol.~113, p.~130503, 2014.

\bibitem{Benedetti_2019}
M.~Benedetti, E.~Lloyd, S.~Sack, and M.~Fiorentini, ``Parameterized quantum circuits as machine learning models,'' {\em Quantum Sci. Technol.}, vol.~5, p.~019601, 2020.

\bibitem{Cerezo_2021_vqa}
M.~Cerezo, A.~Arrasmith, R.~Babbush, S.~C. Benjamin, S.~Endo, K.~Fujii, J.~R. McClean, K.~Mitarai, X.~Yuan, L.~Cincio, and P.~J. Coles, ``Variational quantum algorithms,'' {\em Nat. Rev. Phys.}, vol.~3, pp.~625--644, 2021.

\bibitem{Abbas_2021}
A.~Abbas, D.~Sutter, C.~Zoufal, A.~Lucchi, A.~Figalli, and S.~Woerner, ``The power of quantum neural networks,'' {\em Nat. Comput. Sci.}, vol.~1, pp.~403--409, 2021.

\bibitem{MOYA1996463}
M.~M. Moya and D.~R. Hush, ``Network constraints and multi-objective optimization for one-class classification,'' {\em Neural Networks}, vol.~9, pp.~463--474, 1996.

\bibitem{Khan_2014}
S.~S. Khan and M.~G. Madden, ``One-class classification: taxonomy of study and review of techniques,'' {\em The Knowledge Engineering Review}, vol.~29, p.~345–374, 2014.

\bibitem{perera2021oneclassclassificationsurvey}
P.~Perera, P.~Oza, and V.~M. Patel, ``One-class classification: a survey,'' 2021.

\bibitem{chalapathy2019deeplearninganomalydetection}
R.~Chalapathy and S.~Chawla, ``Deep learning for anomaly detection: a survey,'' 2019.

\bibitem{NIPS1999_8725fb77}
B.~Sch{\"o}lkopf, R.~C. Williamson, A.~J. Smola, J.~Shawe-Taylor, and J.~Platt, ``Support vector method for novelty detection,'' in {\em Advances in Neural Information Processing Systems}, vol.~12, 1999.

\bibitem{Tax2004}
D.~M. Tax and R.~P. Duin, ``Support vector data description,'' {\em Machine Learning}, vol.~54, pp.~45--66, 2004.

\bibitem{pmlr-v80-ruff18a}
L.~Ruff, R.~Vandermeulen, N.~Goernitz, L.~Deecke, S.~A. Siddiqui, A.~Binder, E.~M{\"u}ller, and M.~Kloft, ``Deep one-class classification,'' in {\em Proceedings of the 35th International Conference on Machine Learning}, vol.~80, pp.~4393--4402, 2018.

\bibitem{phua2010comprehensive}
C.~Phua, V.~Lee, K.~Smith, and R.~Gayler, ``A comprehensive survey of data mining-based fraud detection research,'' 2010.

\bibitem{li2012identifying}
S.-H. Li, D.~C. Yen, W.-H. Lu, and C.~Wang, ``Identifying the signs of fraudulent accounts using data mining techniques,'' {\em Computers in Human Behavior}, vol.~28, pp.~1002--1013, 2012.

\bibitem{jeragh2018combining}
M.~Jeragh and M.~AlSulaimi, ``Combining auto encoders and one class support vectors machine for fraudulant credit card transactions detection,'' in {\em 2018 Second World Conference on Smart Trends in Systems, Security and Sustainability (WorldS4)}, pp.~178--184, 2018.

\bibitem{GARCIATEODORO200918}
P.~Garc\'ia-Teodoro, J.~D\'iaz-Verdejo, G.~Maci\'a-Fern\'andez, and E.~V\'azquez, ``Anomaly-based network intrusion detection: Techniques, systems and challenges,'' {\em Computers \& Security}, vol.~28, pp.~18--28, 2009.

\bibitem{feher2014cell}
K.~Feher, J.~Kirsch, A.~Radbruch, H.-D. Chang, and T.~Kaiser, ``Anomaly-based network intrusion detection: Techniques, systems and challenges,'' {\em Bioinformatics}, vol.~30, pp.~3372--3378, 2014.

\bibitem{min2017deep}
S.~Min, B.~Lee, and S.~Yoon, ``Deep learning in bioinformatics,'' {\em Briefings in Bioinformatics}, vol.~18, pp.~851--869, 2017.

\bibitem{Fraser2022}
K.~Fraser, S.~Homiller, R.~K. Mishra, B.~Ostdiek, and M.~D. Schwartz, ``Deep learning in bioinformatics,'' {\em Journal of High Energy Physics}, vol.~2022, p.~851–869, 2022.

\bibitem{Liu_2018}
N.~Liu and P.~Rebentrost, ``Quantum machine learning for quantum anomaly detection,'' {\em Phys. Rev. A.}, vol.~97, p.~042315, 2018.

\bibitem{Kottmann_2021}
K.~Kottmann, F.~Metz, J.~Fraxanet, and N.~Baldelli, ``Variational quantum anomaly detection: Unsupervised mapping of phase diagrams on a physical quantum computer,'' {\em Phys. Rev. Res.}, vol.~3, p.~043184, 2021.

\bibitem{Sakhnenko_2022}
A.~Sakhnenko, C.~O'Meara, K.~J.~B. Ghosh, C.~B. Mendl, G.~Cortiana, and J.~Bernab{\'e}-Moreno, ``Hybrid classical-quantum autoencoder for anomaly detection,'' {\em Quantum Machine Intelligence}, vol.~4, no.~27, 2022.

\bibitem{Ngairangbam_2022}
V.~S. Ngairangbam, M.~Spannowsky, and M.~Takeuchi, ``Anomaly detection in high-energy physics using a quantum autoencoder,'' {\em Phys. Rev. D.}, vol.~105, p.~095004, 2022.

\bibitem{Park_2023}
G.~Park, J.~Huh, and D.~K. Park, ``Variational quantum one-class classifier,'' {\em Machine Learning: Science and Technology}, vol.~4, p.~015006, 2022.

\bibitem{Oh_2024}
H.~Oh and D.~K. Park, ``Quantum support vector data description for anomaly detection,'' {\em Machine Learning: Science and Technology}, vol.~5, p.~035052, 2024.

\bibitem{cong_QCNN}
I.~Cong, S.~Choi, and M.~D. Lukin, ``Quantum convolutional neural networks,'' {\em Nature Physics}, vol.~15, p.~1273–1278, 2019.

\bibitem{726791}
Y.~Lecun, L.~Bottou, Y.~Bengio, and P.~Haffner, ``Gradient-based learning applied to document recognition,'' {\em Proceedings of the IEEE}, vol.~86, pp.~2278--2324, 1998.

\bibitem{xiao2017fashionmnist}
I.~Loshchilov and F.~Hutter, ``Fashion-mnist: a novel image dataset for benchmarking machine learning algorithms,'' 2017.

\bibitem{kagglecredit}
``Kaggle credit card fraud detection: Anonymized credit card transactions labeled as fraudulent or genuine.'' Kaggle.
\newblock Last accessed 4 April 2025.

\bibitem{kagglenetwork}
``Kaggle network intrusion dataset(cic-ids-2017): Anomaly detection in network dataset.'' Kaggle.
\newblock Last accessed 10 September 2025.

\bibitem{bergholm2020pennylane}
V.~Bergholm, J.~Izaac, M.~Schuld, C.~Gogolin, S.~Ahmed, V.~Ajith, M.~S. Alam, G.~Alonso-Linaje, B.~AkashNarayanan, A.~Asadi, J.~M. Arrazola, U.~Azad, S.~Banning, C.~Blank, T.~R. Bromley, B.~A. Cordier, J.~Ceroni, A.~Delgado, O.~D. Matteo, A.~Dusko, T.~Garg, D.~Guala, A.~Hayes, R.~Hill, A.~Ijaz, T.~Isacsson, D.~Ittah, S.~Jahangiri, P.~Jain, E.~Jiang, A.~Khandelwal, K.~Kottmann, R.~A. Lang, C.~Lee, T.~Loke, A.~Lowe, K.~McKiernan, J.~J. Meyer, J.~A. Monta{\~n}ez-Barrera, R.~Moyard, Z.~Niu, L.~J. O'Riordan, S.~Oud, A.~Panigrahi, C.-Y. Park, D.~Polatajko, N.~Quesada, C.~Roberts, N.~S{\'a}, I.~Schoch, B.~Shi, S.~Shu, S.~Sim, A.~Singh, I.~Strandberg, J.~Soni, A.~Sz{\'a}va, S.~Thabet, R.~A. Vargas-Hernández, T.~Vincent, N.~Vitucci, M.~Weber, D.~Wierichs, R.~Wiersema, M.~Willmann, V.~Wong, S.~Zhang, and N.~Killoran, ``Pennylane: Automatic differentiation of hybrid quantum-classical computations,'' 2022.

\bibitem{Bravo_Prieto_2021}
C.~Bravo-Prieto, ``Quantum autoencoders with enhanced data encoding,'' {\em Machine Learning: Science and Technology}, vol.~2, p.~035028, 2021.

\bibitem{hur_NQE}
T.~Hur, I.~F. Araujo, and D.~K. Park, ``Neural quantum embedding: Pushing the limits of quantum supervised learning,'' {\em Phys. Rev. A}, vol.~110, no.~022411, 2024.

\bibitem{hur_margin}
T.~Hur and D.~K. Park, ``Understanding generalization in quantum machine learning with margins,'' 2024.

\bibitem{Havlicek_2019}
V.~Havl{\'i}{\v{c}}ek, A.~D. C{\'o}rcoles, K.~Temme, A.~W. Harrow, A.~Kandala, J.~M. Chow, and J.~M. Gambetta, ``Supervised learning with quantum-enhanced feature spaces,'' {\em Nature}, vol.~567, p.~209–212, 2019.

\bibitem{liu_2025}
H.~Liu, T.~Hur, S.~Zhang, L.~Che, X.~Long, X.~Wang, K.~Huang, Y.~Fan, Y.~Zheng, Y.~Feng, X.~Nie, D.~K. Park, and D.~Lu, ``Neural quantum embedding via deterministic quantum computation with one qubit,'' {\em Phys. Rev. Lett.}, vol.~135, p.~080603, 2025.

\bibitem{shepherd_2009}
D.~Shepherd and M.~J. Bremner, ``Temporally unstructured quantum computation,'' {\em Proc. A}, vol.~465, pp.~1413--1439, 2009.

\bibitem{bremner_2011}
M.~J. Bremner, R.~Jozsa, and D.~J. Shepherd, ``Classical simulation of commuting quantum computations implies collapse of the polynomial hierarchy,'' {\em Proceedings of the Royal Society A: Mathematical, Physical and Engineering Sciences}, vol.~467, p.~459–472, 2010.

\bibitem{bremner_2016}
M.~J. Bremner, A.~Montanaro, and D.~J. Shepherd, ``Average-case complexity versus approximate simulation of commuting quantum computations,'' {\em Phys. Rev. Lett.}, vol.~117, p.~080501, 2016.

\bibitem{PerezSalinas2020datareuploading}
A.~P{\'e}rez-Salinas, A.~Cervera-Lierta, E.~Gil-Fuster, and J.~I. Latorre, ``Data re-uploading for a universal quantum classifier,'' {\em Quantum}, vol.~4, p.~226, 2020.

\bibitem{McClean_2018}
J.~R. McClean, S.~Boixo, V.~N. Smelyanskiy, R.~Babbush, and H.~Neven, ``Barren plateaus in quantum neural network training landscapes,'' {\em Nature Communications}, vol.~9, no.~4812, 2018.

\bibitem{Sim_2019}
S.~Sim, P.~D. Johnson, and A.~Aspuru‐Guzik, ``Expressibility and entangling capability of parameterized quantum circuits for hybrid quantum‐classical algorithms,'' {\em Advanced Quantum Technologies}, vol.~2, no.~12, 2019.

\bibitem{Cerezo_2021_barren_plateaus}
M.~Cerezo, A.~Sone, T.~Volkoff, L.~Cincio, and P.~J. Coles, ``Cost function dependent barren plateaus in shallow parametrized quantum circuits,'' {\em Nature Communications}, vol.~12, no.~1791, 2021.

\bibitem{Wang_2021}
S.~Wang, E.~Fontana, M.~Cerezo, K.~Sharma, A.~Sone, L.~Cincio, and P.~J. Coles, ``Noise-induced barren plateaus in variational quantum algorithms,'' {\em Nature Communications}, vol.~12, no.~6961, 2021.

\bibitem{Grant_2018}
E.~Grant, M.~Benedetti, S.~Cao, A.~Hallam, J.~Lockhart, V.~Stojevic, A.~G. Green, and S.~Severini, ``Hierarchical quantum classifiers,'' {\em npj Quantum Information}, vol.~4, no.~65, 2018.

\bibitem{Huang_2020}
H.-Y. Huang, R.~Kueng, and J.~Preskill, ``Predicting many properties of a quantum system from very few measurements,'' {\em Nature Physics}, vol.~16, pp.~1050--1057, 2020.

\bibitem{pesah2020absence}
A.~Pesah, M.~Cerezo, S.~Wang, T.~Volkoff, A.~T. Sornborger, and P.~J. Coles, ``Absence of barren plateaus in quantum convolutional neural networks,'' {\em Phys. Rev. X.}, vol.~11, p.~041011, 2021.

\bibitem{Banchi_2021}
L.~Banchi, J.~Pereira, and S.~Pirandola, ``Generalization in quantum machine learning: A quantum information standpoint,'' {\em PRX Quantum}, vol.~2, p.~040321, 2021.

\bibitem{hur_QCNN}
T.~Hur, L.~Kim, and D.~K. Park, ``Quantum convolutional neural network for classical data classification,'' {\em Quantum Machine Intelligence}, vol.~4, no.~1, 2022.

\bibitem{kim2023classical}
J.~Kim, J.~Huh, and D.~K. Park, ``Classical-to-quantum convolutional neural network transfer learning,'' {\em Neurocomputing}, vol.~555, p.~126643, 2023.

\bibitem{liu2017survey}
W.~Liu, Z.~Wang, X.~Liu, N.~Zeng, Y.~Liu, and F.~E. Alsaadi, ``A survey of deep neural network architectures and their applications,'' {\em Neurocomputing}, vol.~234, pp.~11--26, 2017.

\bibitem{Verteletskyi_2020}
V.~Verteletskyi, T.-C. Yen, and A.~F. Izmaylov, ``Measurement optimization in the variational quantum eigensolver using a minimum clique cover,'' {\em The Journal of Chemical Physics}, vol.~152, p.~124114, 2020.

\bibitem{Schuld_2019}
M.~Schuld, V.~Bergholm, C.~Gogolin, J.~Izaac, and N.~Killoran, ``Evaluating analytic gradients on quantum hardware,'' {\em Phys. Rev. A.}, vol.~99, p.~032331, 2019.

\bibitem{kingma2017adammethodstochasticoptimization}
D.~P. Kingma and J.~Ba, ``Adam: a method for stochastic optimization,'' 2017.

\bibitem{loshchilov2017sgdrstochasticgradientdescent}
I.~Loshchilov and F.~Hutter, ``Sgdr: stochastic gradient descent with warm restarts,'' 2017.

\bibitem{Bravyi_2021}
S.~Bravyi, S.~Sheldon, A.~Kandala, D.~C. McKay, and J.~M. Gambetta, ``Mitigating measurement errors in multiqubit experiments,'' {\em Phys. Rev. A.}, vol.~103, p.~042605, 2021.

\bibitem{Nation_2021}
P.~D. Nation, H.~Kang, N.~Sundaresan, and J.~M. Gambetta, ``Scalable mitigation of measurement errors on quantum computers,'' {\em PRX Quantum}, vol.~2, p.~040326, 2021.

\bibitem{Kwon_2021}
H.~Kwon and J.~Bae, ``A hybrid quantum-classical approach to mitigating measurement errors in quantum algorithms,'' {\em IEEE Transactions on Computers}, vol.~70, p.~1401–1411, 2021.

\bibitem{van_den_Berg_2022}
E.~v.~d. Berg, Z.~K. Minev, and K.~Temme, ``Model-free readout-error mitigation for quantum expectation values,'' {\em Phys. Rev. A}, vol.~105, p.~032620, 2022.

\bibitem{Kim_2022}
J.~Kim, B.~Oh, Y.~Chong, E.~Hwang, and D.~K. Park, ``Quantum readout error mitigation via deep learning,'' {\em New Journal of Physics}, vol.~24, p.~073009, 2022.

\bibitem{Lee_2023}
C.~Lee and D.~K. Park, ``Scalable quantum measurement error mitigation via conditional independence and transfer learning,'' {\em Machine Learning: Science and Technology}, vol.~4, p.~045051, 2023.

\bibitem{Bonet_Monroig_2023}
X.~Bonet-Monroig, H.~Wang, D.~Vermetten, B.~Senjean, C.~Moussa, T.~B{\"a}ck, V.~Dunjko, and T.~E. O'Brien, ``Performance comparison of optimization methods on variational quantum algorithms,'' {\em Phys. Rev. A.}, vol.~107, p.~032407, 2023.

\end{thebibliography}

\appendix

\setcounter{section}{0}
\setcounter{figure}{0}
\setcounter{table}{0}
\setcounter{equation}{0}
\setcounter{algorithm}{0}

\renewcommand{\thesection}{\Alph{section}} 
\renewcommand\thefigure{\Alph{section}.\arabic{figure}} 
\renewcommand\theequation{\Alph{section}.\arabic{equation}} 
\renewcommand\thetable{\Alph{section}.\arabic{table}} 
\renewcommand\thealgorithm{\Alph{section}.\arabic{algorithm}}

{
\vspace{0.5em}
\flushleft
  \LARGE\bfseries Appendix\par
  \vspace{0.5em}
}

\section{NQSVDD simulation results}
\label{appendix}
\begin{figure}[ht]
    \centering
    \subfloat[MNIST "0"\label{fig:bylatent_mnist}]{
    \includegraphics[width=3.0in]{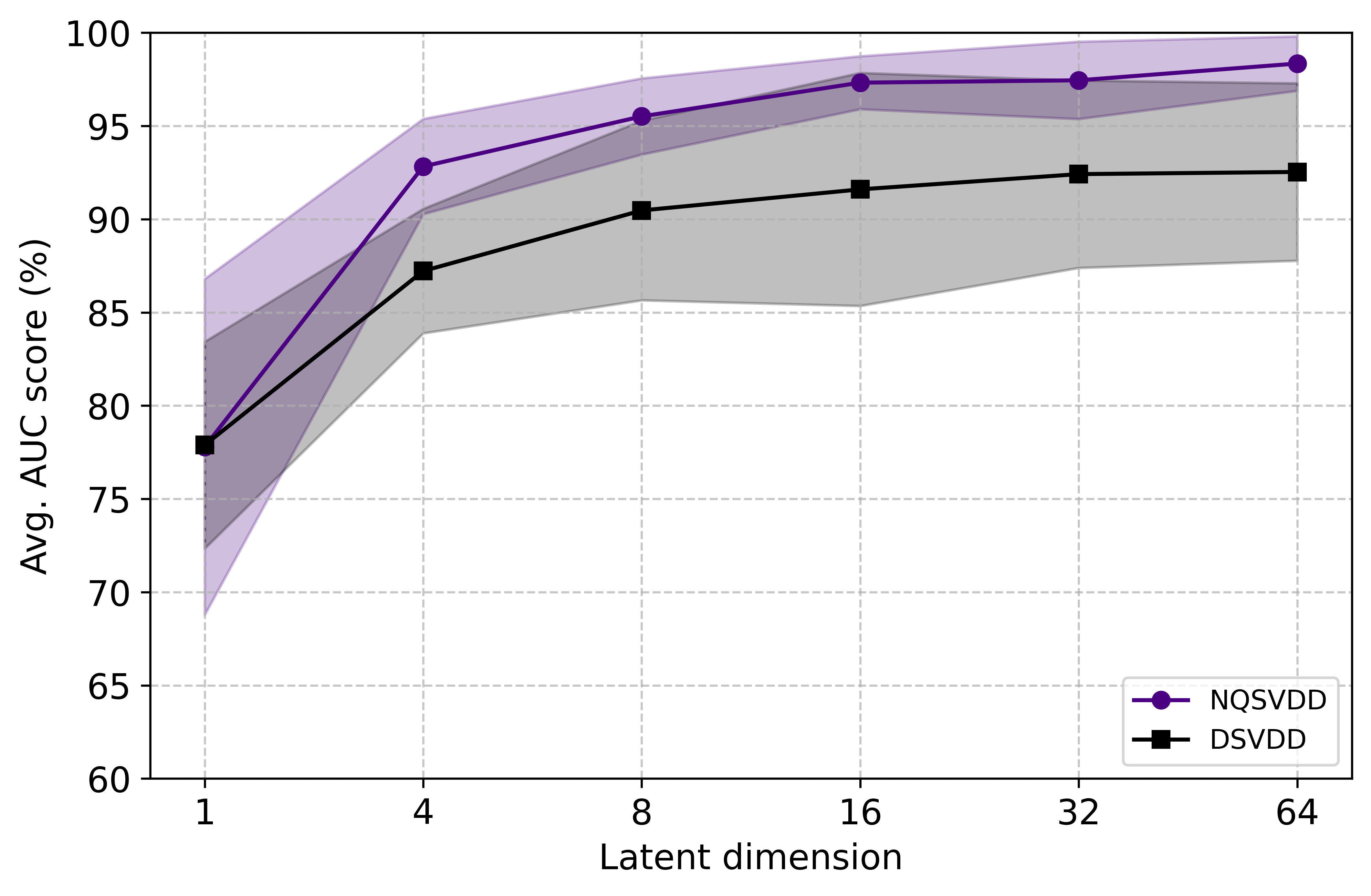}
    }
    \hfill
    \subfloat[Fashion-MNIST "T-shirt"\label{fig:bylatent_fashion}]{
    \includegraphics[width=3.0in]{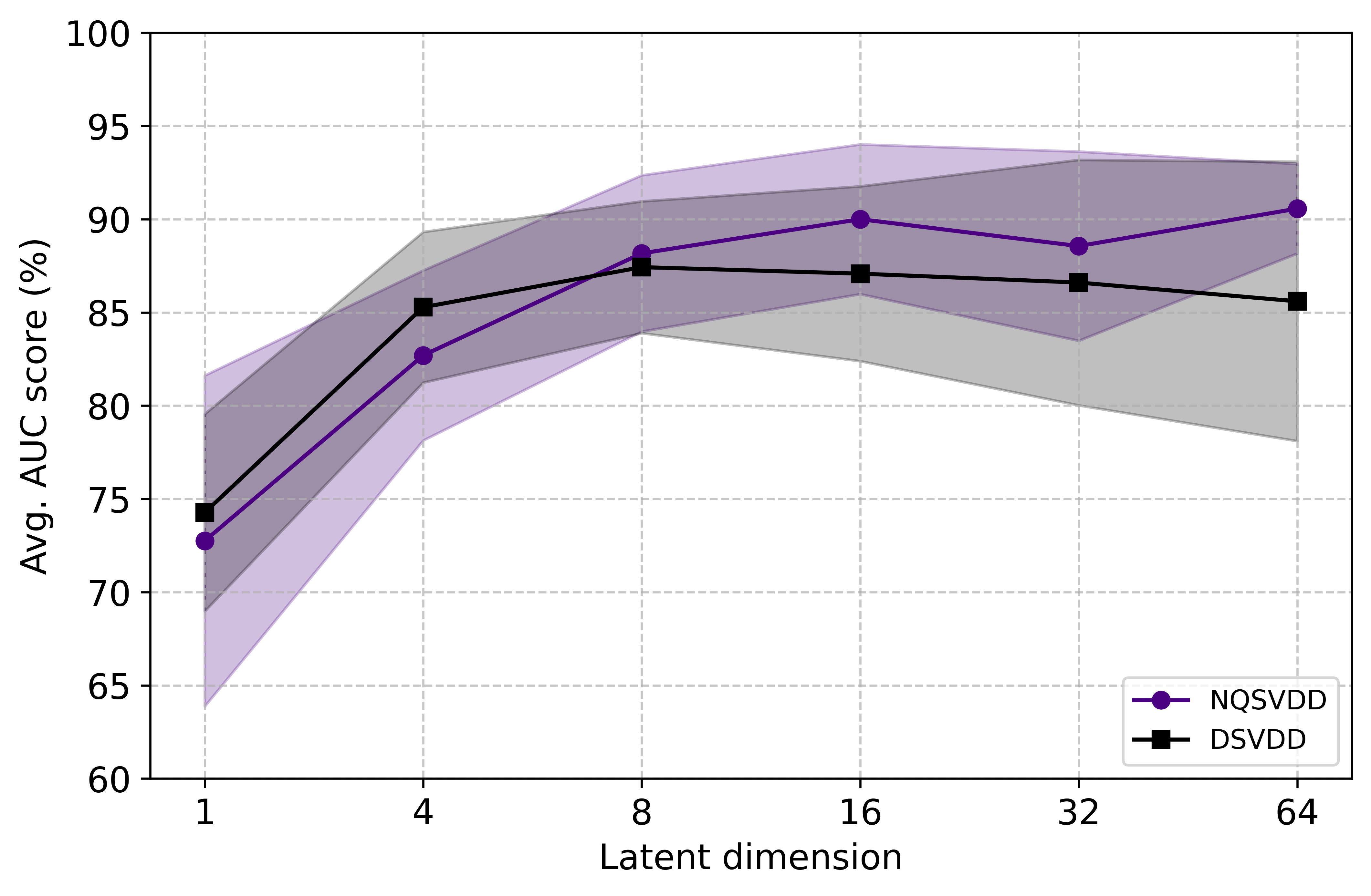}
    }
    \caption{
    Avg. AUC (\%) and std. (error bar) by the dimension of latent space on (a) ``0" of MNIST and (b) ``T-shirt" of Fashion-MNIST datasets with $10$ random seeds per latent dimension.
    }
\end{figure}

\begin{figure}[ht]
    \centering
    \includegraphics[width=3.0in]{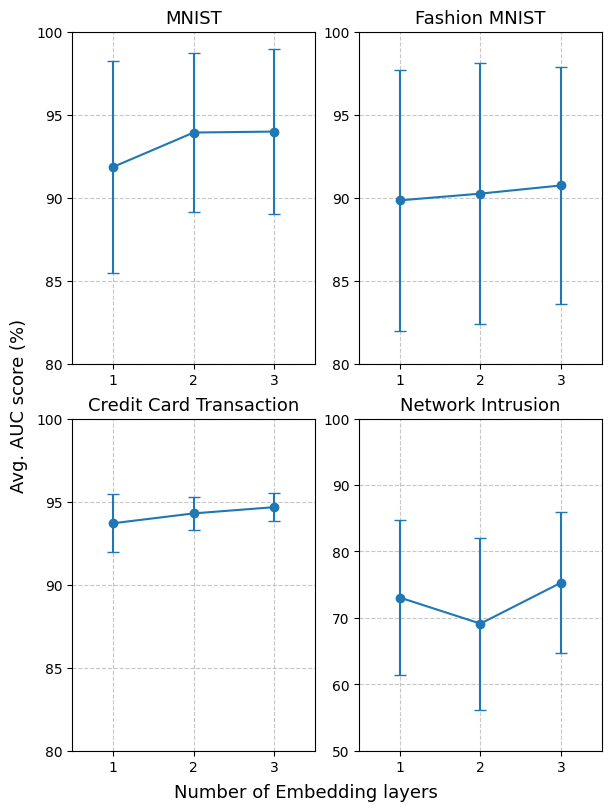}
    \caption{Avg. AUC (\%) and std. (error bar) for different numbers of embedding layers in NQSVDD. For MNIST and Fashion MNIST, the average and std. of AUC scores are computed over all target-class experiments, using 10 random seeds per target class (100 experiments per the number of embedding layers). For Credit Card Transaction and Network Intrusion datasets, results are averaged over 10 random seeds for each number of embedding layers.}
    \label{fig:byemb}
\end{figure}

\begin{figure}[ht]
    \centering
    \subfloat[Iteration \(0\)\label{fig:visualization_iter0}]{
    \includegraphics[width=3.0in]{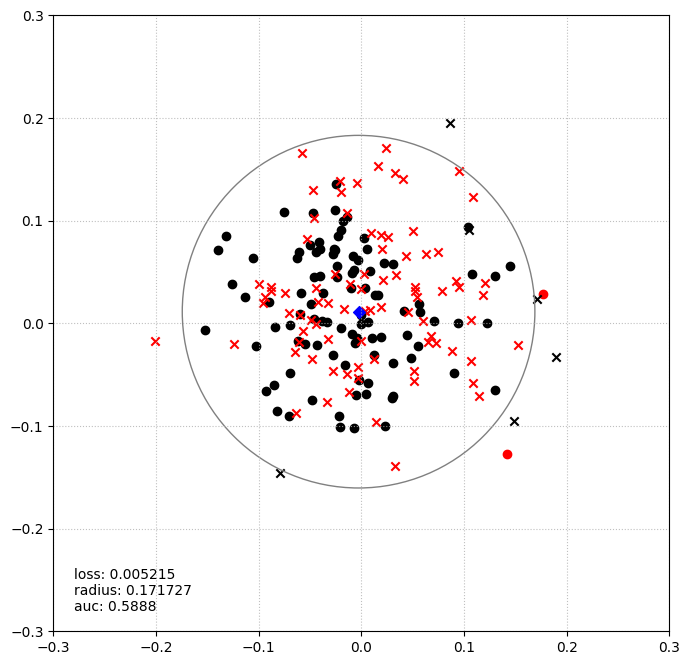}
    }
    \hfill
    \subfloat[Iteration \(1500\)\label{fig:visualization_iter1500}]{
    \includegraphics[width=3.0in]{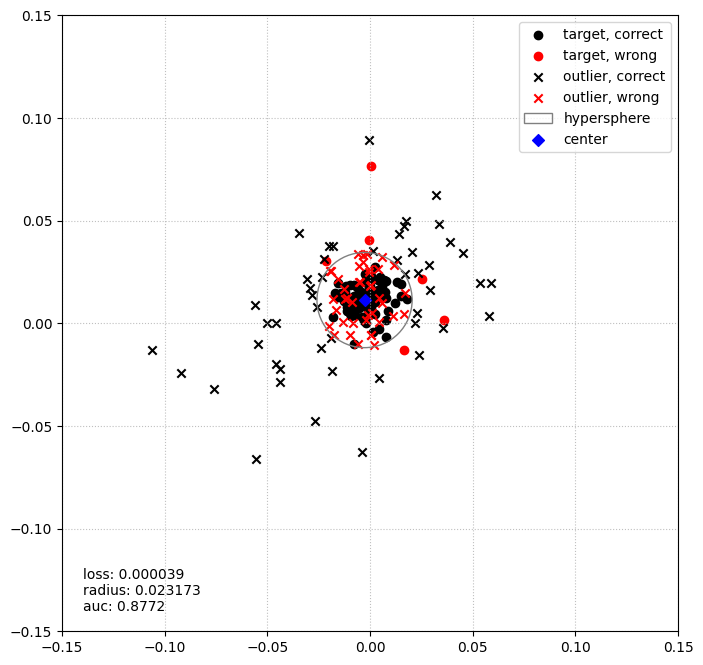}
    }
    \caption{
    Visualization of NQSVDD outputs for MNIST test data in a two-dimensional space. To enable visualization, the output dimension of NQSVDD is set to two. The upper plot illustrates the output of the initial model, and the bottom plot shows the mapping after $1500$ training iterations.
    }
    \label{fig:visualization_mnist0}
\end{figure}

\begin{figure}[ht]
    \centering
    \subfloat[Iteration \(0\)\label{fig:fashion0_iter0}]{
    \includegraphics[width=3.0in]{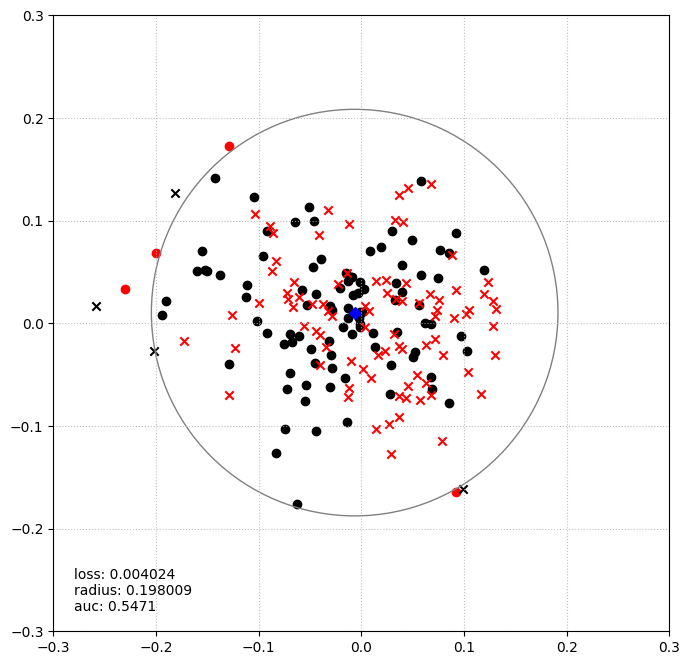}
    }
    \hfill
    \subfloat[Iteration \(1500\)\label{fig:fashion0_iter1500}]{
    \includegraphics[width=3.0in]{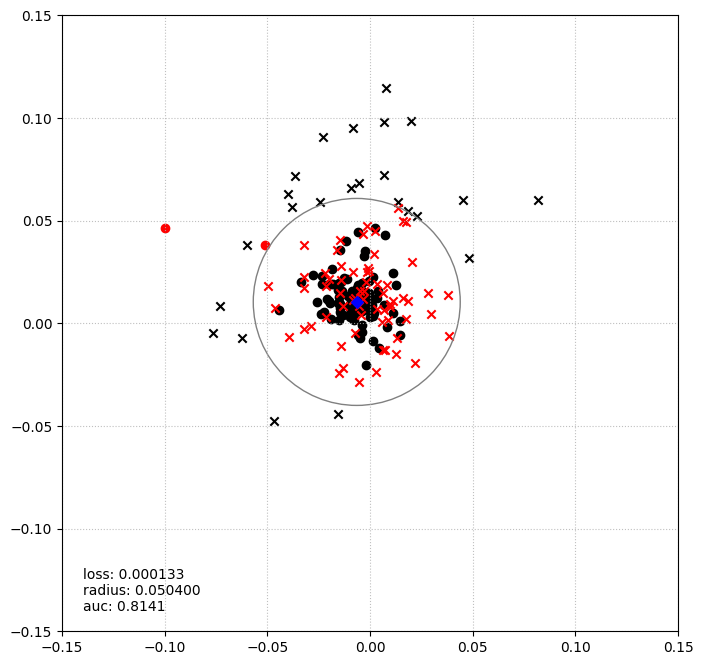}
    }
    \caption{
    Visualization of NQSVDD outputs for Fashion-MNIST test data in a two-dimensional space. To enable visualization, the output dimension of NQSVDD is set to two. The upper plot illustrates the output of the initial model, and the bottom plot shows the mapping after $1500$ training iterations.
    }
    \label{fig:visualization_fashion0}
\end{figure}

\noindent In this section, we provide additional results of NQSVDD simulation. Figures \ref{fig:bylatent_mnist}, \ref{fig:bylatent_fashion} and \ref{fig:byemb} examine the OCC performance as the model scale increases, in terms of the latent space dimension and the number of embedding layers. As shown in Fig. \ref{fig:bylatent_mnist} and Fig. \ref{fig:bylatent_fashion}, increasing the latent space improves performance for both NQSVDD and DSVDD methods, while NQSVDD consistently achieves higher AUC scores across all dimensions. Figure \ref{fig:byemb} shows that increasing the number of embedding layers in NQSVDD yields modest performance improvements across datasets. Figure \ref{fig:visualization_mnist0} and \ref{fig:visualization_fashion0} illustrate the effectiveness of NQSVDD feature mapping in building a compact, minimum-volume hypersphere that encloses the target data points after training. 

\end{document}